\documentclass[a4paper,aps,pra,reprint,onecolumn,superscriptaddress,showpacs,nofootinbib,notitlepage,noeprint]{revtex4-1}

\usepackage{amsbsy}
\usepackage{amsfonts}
\usepackage{amsmath}
\usepackage{amssymb}
\usepackage{anyfontsize}
\usepackage{array}
\usepackage{bbding}
\usepackage{bbm}
\usepackage{bm}
\usepackage{color}
\usepackage{curves}
\usepackage{epsfig}
\usepackage{fixmath}
\usepackage[T1]{fontenc}
\usepackage[margin=1in,bottom=1.14in]{geometry}
\usepackage{graphicx}
\usepackage{multirow}
\usepackage{placeins}
\usepackage[svgnames]{xcolor}
\usepackage{tikz}
\usetikzlibrary{svg.path}
\usepackage{hyperref}
\hypersetup{colorlinks=true,allcolors=blue}
\usepackage{relsize}
\usepackage{rotating}
\usepackage{scalerel}
\usepackage{txfonts}
\usepackage{ulem}
\usepackage{url}

\definecolor{orcidlogocol}{HTML}{A6CE39}
\tikzset{
  orcidlogo/.pic={
    \fill[orcidlogocol] svg{M256,128c0,70.7-57.3,128-128,128C57.3,256,0,198.7,0,128C0,57.3,57.3,0,128,0C198.7,0,256,57.3,256,128z};
    \fill[white] svg{M86.3,186.2H70.9V79.1h15.4v48.4V186.2z}
                 svg{M108.9,79.1h41.6c39.6,0,57,28.3,57,53.6c0,27.5-21.5,53.6-56.8,53.6h-41.8V79.1z M124.3,172.4h24.5c34.9,0,42.9-26.5,42.9-39.7c0-21.5-13.7-39.7-43.7-39.7h-23.7V172.4z}
                 svg{M88.7,56.8c0,5.5-4.5,10.1-10.1,10.1c-5.6,0-10.1-4.6-10.1-10.1c0-5.6,4.5-10.1,10.1-10.1C84.2,46.7,88.7,51.3,88.7,56.8z};
  }
}

\newcommand\orcid[1]{\href{https://orcid.org/#1}{\mbox{\scalerel*{
\begin{tikzpicture}[yscale=-1,transform shape]
\pic{orcidlogo};
\end{tikzpicture}
}{|}}}}

\definecolor{darkgreen}{RGB}{30,150,30}
\definecolor{darkblue}{RGB}{0,0,130}
\definecolor{darkred}{rgb}{.8,0,0}

\DeclareMathAlphabet{\mathantt}{OT1}{antt}{li}{it}
\DeclareMathAlphabet{\mathpzc}{OT1}{pzc}{m}{it}

\newcommand{\negphantom}[1]{\ifmmode\settowidth{\dimen0}{$#1$}\else\settowidth{\dimen0}{#1}\fi\hspace*{-\dimen0}}
\newcommand{\tr}{\mathrm{tr}}

\renewcommand{\vec}[1]{\mathbold{#1}}
\newcommand{\nn}{\nonumber}
\renewcommand{\d}{\mathrm{d}}

\newcommand{\bok}[3]{\left<\right.\hspace{-0.5ex}{#1}\left.\hspace{-0.5ex}\right|{#2}\left|\right.\hspace{-0.5ex}{#3}\left.\hspace{-0.5ex}\right>}

\newcommand{\mw}[1]{\left<\right.\hspace{-0.5ex}{#1}\left.\hspace{-0.5ex}\right>}

\newcommand{\nab}{\boldsymbol{\nabla}}

\newcommand{\W}{_{\mathrm{W}}}
\newcommand{\e}[1]{\mathrm{e}^{#1}}

\newcommand{\Int}[1][-5pt]{\int\limits_{\begin{picture}(16,3)(-8,-3)%
		\put(0,0){\curve(-3,0,-8,0)\curve(3,0,8,0)}%
		\put(8,0){\curve(0,0,-1.5,1.5)\curve(0,0,-1.5,-1.5)}%
		\put(0,0){\arc(-3,0){180}}\put(0,0){\makebox(0,0){$\cdot$}}%
		\end{picture}}\hspace*{#1}}

\definecolor{ao(english)}{rgb}{0.0, 0.5, 0.0}

\newcommand{\ignore}[1]{\relax}

\newcommand{\I}{\mathrm{i}}

\DeclareMathAlphabet{\vecfont}{T1}{cmr}{bx}{it}
\DeclareMathAlphabet{\nfont}{T1}{cmr}{bx}{sl}

\DeclareMathAlphabet{\cfont}{T1}{cmss}{sbc}{n}

\renewcommand{\emph}{\textit}

\definecolor{RoyalBlue}{cmyk}{1, 0.50, 0, 0}

\DeclareMathAlphabet{\mathpzc}{OT1}{pzc}{m}{it}

\begin{document}

\fontsize{11}{12}\selectfont

\title{Atoms, dimers, and nanoparticles from orbital-free density--potential functional theory}

\author{Martin-Isbj\"orn~\surname{Trappe}\orcid{0000-0002-2911-4162}}
\thanks{\href{mailto:martin.trappe@quantumlah.org}{martin.trappe@quantumlah.org}; corresponding author}
\affiliation{Centre for Quantum Technologies, National University of Singapore, 3 Science Drive 2, Singapore 117543, Singapore}

\author{William~C.~\surname{Witt}\orcid{0000-0002-1578-1888}}
\thanks{\href{mailto:wcw28@cam.ac.uk}{wcw28@cam.ac.uk}}
\affiliation{Department of Materials Science \& Metallurgy, University of Cambridge, 27 Charles Babbage Road, Cambridge, CB3 0FS} 
  
\author{Sergei~\surname{Manzhos}\orcid{0000-0001-8172-7903}}
\thanks{\href{mailto:manzhos.s.aa@m.titech.ac.jp}{manzhos.s.aa@m.titech.ac.jp}}
\affiliation{School of Materials and Chemical Technology, Tokyo Institute of Technology, Ookayama 2-12-1, Meguro-ku, Tokyo 152-8552, Japan}

\date{\today}

\begin{abstract}
Density--potential functional theory (DPFT) is an alternative formulation of orbital-free density functional theory that may be suitable for modeling the electronic structure of large systems. To date, DPFT has been applied mainly to quantum gases in one- and two dimensional settings. In this work, we study the performance of DPFT when applied to real-life systems: atoms, dimers, and nanoparticles. We build on systematic Suzuki--Trotter factorizations of the quantum-mechanical propagator and on the Wigner function formalism, respectively, to derive nonlocal as well as semilocal functional approximations in complete analogy to their well-established lower-dimensional versions---without resorting to system-specific approximations or ad-hoc measures of any kind. The cost for computing the associated semiclassical ground-state single-particle density scales (quasi-)linearly with particle number. We illustrate that the developed density formulae become relatively more accurate for larger particle numbers, can be improved systematically, are quite universally applicable, and, hence, may offer alternatives to existing orbital-free methods for mesoscopic quantum systems.
\end{abstract}

\maketitle


\section{\label{Introduction}Introduction}

Electronic structure calculations form the theoretical basis of today's materials science and chemistry, where demand for reliable predictions of electronic energies and electron distributions has been growing steadily for decades \cite{Hohenberg1964,Kohn1965,Burke2012,Becke2014,Hasnip2014,Kalita2021}. The majority of applications rely on Kohn--Sham density functional theory (KS--DFT) \cite{Kohn1965}, which self-consistently builds the spatial electron density distribution $n$ and the associated energy $E[n]$ from orbitals of an auxiliary, effectively noninteracting system. Since these orbitals have to be mutually orthogonal, the computational cost of standard KS--DFT scales cubically with the electron number $N$. Consequently, high-throughput KS calculations are typically limited to a few hundred atoms, although linear-scaling implementations can target much larger particle numbers, commonly achieved through localized orbitals and massive parallelization \cite{Soler2002,Bowler2012,Mohr2015,Cole2016,Nakata2020,Prentice2020}. However, these computationally efficient approaches rely on known or presupposed properties of the target systems and are not easily transferable beyond the class of systems they are designed for \cite{Aarons2016}.

A trade-off among scalability, accuracy, and transferability is unavoidable. With emphasis on efficiency for large $N$, orbital-free density functional theory (OF--DFT) naturally becomes the electronic structure method of choice for routinely investigating thousands to millions of interacting particles \cite{Chen2016,Witt2018}, without relying on particular approximations from the outset like, for example, in the density functional tight binding method \cite{Elstner2014}. OF--DFT is the original (Hohenberg--Kohn) formulation of DFT, which has been in the shadows for decades due to the absence of accurate enough functionals for the kinetic energy. Today, however, the field of OF-DFT is enjoying rapid growth due to a constellation of factors that include advanced computational tools, such as modern optimizers and machine learning \cite{Golub2019,Fujinami2020,Manzhos2020,Lueder2020,Tan2023}, as well as new theoretical developments \cite{Witt2019a,Witt2019b,Witt2019c,Shao2021b,Jiang2021,Jiang2022,DellaSala2022,Xu2022}.

OF--DFT primarily aims at utilizing efficient and accurate functionals for the kinetic energy $E_{\mathrm{kin}}$, since (i) other major contributions to the total energy, such as the electron--ion or Hartree energy, are known and simple density functionals, and (ii) a generic exchange--correlation energy functional such as PBE \cite{Perdew1996,Perdew1997Erratum} can be accurate enough for many OF--DFT studies of electronic systems in three dimensions (3D). Importantly, we note that DFT implementations are less developed for lower-dimensional settings and non-coulombic interactions, not least due to the focus on 3D electronic systems ultimately required for chemistry and materials applications. For example, exchange--correlation functionals for 2D electronic systems have been rigorously derived only in recent history \cite{Vilhena2014}. Moreover, some highly accurate approximations cannot be extended beyond 1D geometries \cite{Ribeiro2015,Okun2023DFMPS}, and the interactions in ultracold Fermi gases demands DFT techniques that deviate markedly from traditional routes \cite{Ma2012,Trappe2016,Trappe2017,Trappe2021b}. While functionals for 3D are needed in the vast majority of use cases, we speculate that the most transferable functionals will extend seamlessly into low-dimensional settings, and that the development of functionals for 3D will benefit from lessons learned with tractable systems in 1D and 2D.

Much of scholarly material on OF--DFT features approximations of the kinetic energy \textit{density} functional $E_{\mathrm{kin}}[n]$ (KEDF), for which even the general noninteracting version is not known explicitly (in position--space representation). The following selection of OF--DFT approaches provides an overview pertinent to the present work, but is by no means complete; see Refs.~\cite{Witt2018,Xu2020} and references therein for broader surveys. Practical implementations of OF--DFT are accessible through software packages like GPAW \cite{Lehtomaeki2014}, PROFESS \cite{Chen2015,Dieterich2017b}, ATLAS \cite{Mi2016,Shao2018,Xu2020}, CONUNDrum \cite{Golub2020}, and DFTpy \cite{Shao2021b}, which build on nonlocal and semilocal KEDFs that prove appropriate for certain classes of systems. In contrast to rudimentary approximations such as the prototypical Thomas--Fermi (TF) model, nonlocal KEDFs are capable of producing the typical oscillations of quantum-mechanical densities that can be viewed as aggregations of many KS-type single-particle densities. But these accurate functionals are limited mainly to main group metals with inroads into bulk semiconductors \cite{Wang1992,Wang1998,Wang1999,Huang2010}. Moreover, their application commonly depends on the average density in the unit cell, such that addressing isolated systems remains problematic \cite{Constantin2019}, although efforts are made to overcome these shortcomings, for example, through a KEDF \cite{Xu2020} that is inspired by the local density approximation (LDA). Alternatively, rigorous expansions in terms of the density gradient have been considered \cite{Murphy1981,Yang1986,Dreizler1990}, but technical difficulties in adequately resumming higher-order terms of this asymptotic series for obtaining systematic and, hence, maximally transferable KEDFs have impeded widespread application, though very accurate gradient expansions can be constructed for special cases in 1D \cite{Sergeev2016}. The truncation of the gradient expansion renders corresponding KEDFs valid only in regions of slowly varying density---a feature shared with modern semilocal KEDFs that incorporate the gradient and Laplacian of the density \cite{Luo2018,Constantin2018,Constantin2019,Constantin2021}. But some of these functionals, which specifically target electronic systems in 3D, are of limited use for regions of highly variable density. Furthermore, gradient-expanded KEDFs that improve upon the TF and von-Weizs\"acker (vW) KEDFs \cite{Dreizler1990} are commonly (i) parameterized in an ad-hoc fashion and fitted to existing data like in \cite{EspinosaLeal2015} or (ii) take into account constraints that only apply to 3D electronic matter \cite{Garcia-Aldea2012,Karasiev2013}. Specific combinations of nonlocal and semilocal terms have been explored, for example, such that a number of exact constraints are met upon optimizing free parameters \cite{Garcia-Aldea2007,Mi2018}. KEDFs can also be constructed directly from models of the one-body reduced density matrix for electronic systems \cite{Chakraborty2017}. Some OF--DFT variants achieve accuracy and efficiency through electron densities restricted to families of functions for the constituting atoms, partitioning of the target system, and other ad-hoc measures. Applications of OF--DFT can also benefit from problem-specific machine-learned functionals, some of which deliver chemical accuracy, though general-purpose functionals have yet to materialize \cite{Yao2016,Golub2019,Fujinami2020,vonLilienfeld2020,Westermayr2021}. The kinetic as well as total energy can also be expressed in terms of the external potential, which is bijectively related to the particle density. In the resulting potential functional theories \cite{Yang2004,Cangi2013}, both the kinetic energy in general and gradient expansions in particular are accessible more naturally than in DFT. However, the according OF--DFT implementations have so far improved upon TF-type models only for selected systems, and extensions to 3D prove difficult \cite{Peng2012,Cangi2013,Elliott2015}.

In summary, the fundamental importance of the quantum many-body problem across scientific branches and the difficulty in finding general-purpose solutions have motivated a multitude of approaches to OF--DFT that predominantly target electronic structure calculations. Though each of these methods presents a viable route to address specific sets of issues, none offers a clear path of \textit{systematic and nonempirical} improvement \textit{toward} the exact energy functional across virtually all fermion systems, that is, across dimensions, type of interactions, number of fermion species, boundary conditions, and for observables in momentum- as well as in position space. Such a universal orbital-free methodology would allow us to transfer lessons learned in studying very disparate systems and would enjoy widespread application if demands on accuracy and computational efficiency are met. The present work features density potential functional theory (DPFT), which promises to deliver such a unified, parameter-free framework for practically relevant applications.

The conception of DPFT by J.~Schwinger and B.-G. Englert dates back to the early 1980s, with a series of works that developed the semiclassical atom \cite{Englert1982,Englert1984,Englert1984b,Englert1984c,Englert1985a,Englert1985b,Englert1988,Cinal1992,Englert1992,Cinal1993}; see Ref.~\cite{Englert2019articleEntry} for a review of these early developments, which relied on the properties of the central nuclear potential and cannot easily be transferred to molecules and materials. The basis of DPFT is an exact bifunctional for the total energy that depends on (i) the particle density $n$ and (ii) an effective potential $V$ that merges the external potential with the interaction effects. The kinetic energy in DPFT is expressed through the Legendre transform of the KEDF. In reformulating the Hohenberg--Kohn energy functional, this procedure yields an explicit expression for the noninteracting kinetic energy functional in terms of $V$, permits systematic approximations, and thereby alleviates the need of searching for approximations of the unknown $E_{\mathrm{kin}}[n]$. DPFT has been applied to (i) noninteracting systems for benchmarking purposes \cite{Trappe2016,Trappe2017,Chau2018,Trappe2023DFMPS}, (ii) systems in one \cite{Trappe2023DFMPS}, two \cite{Trappe2016,Trappe2017,Trappe2019,Trappe2021b}, and three \cite{Chau2018} dimensions, (iii) small \cite{Chau2018,Trappe2023DFMPS,Trappe2021b} and large \cite{Trappe2017,Trappe2019,Trappe2023DFMPS,Trappe2021b} particle numbers, (iv) graphene heterostructures \cite{Trappe2019}, (v) single atoms \cite{Englert1982,Englert1984,Englert1984b,Englert1984c,Englert1985a,Englert1985b,Englert1988,Cinal1992,Englert1992,Cinal1993}, and (vi) interacting Fermi gases \cite{Trappe2016,Hue2020a,Trappe2021b}. The unifying feature of all these applications is the systematic, parameter-free methodology of the DPFT approximations. In fact, the core principles of DPFT have permeated beyond physics: Ref.~\cite{Trappe2023NatComm} offers a unified theory for ecology based on the mathematical structure of DPFT. Besides DPFT implementations aimed at periodic systems, an important missing element in this array of applications is the calculation of electronic structure.

The present work establishes a new avenue for electronic structure calculations of isolated systems based on semiclassical approximations for DPFT. We employ two disjunct schemes that feature nonlocal and semilocal approximations, respectively. First, we (i) utilize the formulae for quantum-corrected densities derived in Ref.~\cite{Chau2018}, where Suzuki--Trotter factorizations of the time-evolution operator produce a hierarchy of systematic improvements upon the TF approximation without a gradient expansion, and (ii) derive the kinetic energy through an accordingly approximated one-body density matrix. In extending this approach to finite temperature, we reduce the computational cost of the quantum-corrected densities to quasi-linear scaling. The second approximation scheme builds on the Wigner function formalism and delivers semilocal, linearly-scaling formulae for densities and energies through `Airy-averaged' gradient expansions that address evanescent regions accurately and efficiently across dimensions. Here, we derive the expressions for densities and energies for 3D systems, in complete analogy to the 1D and 2D settings that are covered in Refs.~\cite{Trappe2021b,Trappe2023DFMPS} and \cite{Trappe2016,Trappe2017,Trappe2021b}, respectively. Our primary objective is the proof-of-principle applicability of these two DPFT approximation schemes to electronic structure problems. We show that even the next-to-leading-order semiclassical DPFT densities improve significantly upon their lowest order (the TF density). Unfortunately, while some modern KS exchange--correlation functionals sacrifice accuracy of densities in favor of accurate total energies \cite{Medvedev2017}, the reverse seems to hold for the DPFT approximations used in this work: We find the improvements of total energies (as well as energy differences, which matter ultimately) over the TF energies to be minor in practice---although the improvements are sufficient to bind H$_2$, an outcome that cannot be obtained from a self-consistent TF calculation. 

This is the first study of applicability of DPFT approximation schemes to real-life electronic structure. The test cases range from single atoms to relatively large (by ab initio standards) nanoparticles. Although the semiclassical DPFT approximations are expected to become relatively accurate only for large particle numbers, we will calculate atomic densities and dimer properties. These preparatory calculations will help establish the reliability of our approximation schemes for mesoscopic nanoparticles. Metal nanoparticles composed of ${\sim10-10^6}$ atoms, are an important active area of research that fuels modern technologies, including photovoltaics, catalysis, and drug delivery \cite{Lan2014,Tyo2015,Attia2017,Jaeger2018,Hui2019,Xu2020}, in particular because the large ratio between surface and bulk atoms in metal nanoparticles yields unique and useful mechanical, chemical, optical, and electrical properties that differ markedly from their solid-state versions. In isolated systems like metal nanoparticles, the electron density easily drops ten orders of magnitude within Angstroms when crossing the evanescent region from the bulk on the way to vacuum. Such inhomogeneities require special care, as discussed, for example, in Ref.~\cite{Mi2019}, where accurate OF--DFT densities and energies are reported based on an LDA-inspired KEDF. Comprehensive KS analyses of more exotic metal nanoparticles have only recently become possible and are commonly combined with advanced global optimization techniques that deliver the low-energy nanoparticle configurations \cite{Do2012,Jaeger2018,Cuny2018}. For example, geometries of a gold-copper nanoalloy composed of just over 300 atoms are found with a genetic algorithm in Ref.~\cite{Lysgaard2015}. Disregarding stringent demands on accuracy and transferability, the issue of computational cost can likely be resolved with modern OF--DFT implementations \cite{Shao2021}: A refined periodic code that can process any KEDFs recently produced the electronic structure of a one-million-atom Al-nanoparticle on a single CPU \cite{Shao2021b}---but, of course, such impressive outcomes inherit the shortcomings of the underlying KEDFs.

This work is organized as follows. In Sec.~\ref{secDPFT} we introduce the method of orbital-free DPFT. Section~\ref{secGeneralformalism} recapitulates the general DPFT formalism, subsequently supplemented with systematic semiclassical approximations that derive from Suzuki--Trotter factorizations (Sec.~\ref{SectionST}) and approximate Wigner functions (Sec.~\ref{SectionAiry}), respectively. In Sec.~\ref{secNuclearBackground} we declare the pseudopotentials we use for valence- and all-electron calculations. In Sec.~\ref{Results}, we present our main results for the electronic structure of atoms (Sec.~\ref{secAtoms}), dimers (Sec.~\ref{secDimers}), and nanoparticles (Sec.~\ref{secNanoparticles}). We conclude in Sec.~\ref{secConclusions} with the wider implications of this work and its potentially fruitful extensions. In the appendices we collect details on our numerical procedures and on the derivations of the approximate DPFT functionals of energy and density for three-dimensional settings as introduced in Secs.~\ref{SectionST} and \ref{SectionAiry}.

\section{\label{secDPFT}Density-potential functional theory for electronic structure}

The computational cost of OF--DFT is not explicitly dependent on the particle number. But in practice, and in particular for electronic structure, an $N$-dependence is introduced via numerical integration grids that have to be large enough for sufficiently converging densities and energies. A moderate grid size can be adequate even for millions of particles, for example, for some interacting ultracold atomic gases, whereas electronic structure calculations demand an adequate sampling of the electron distribution around \textit{each} nucleus, which makes the OF--DFT approach scale with the number of atoms or number $N$ of electrons---with the proviso that the employed KEDFs are efficient enough, comparable to the TF-KEDF in the best case. The latter holds for many semilocal functionals, while nonlocal KEDFs incur costs that scale at least like ${N\,\mathrm{log}N}$. Furthermore, all that holds only if the cost of the interaction functional is small enough---a lower bound is usually put in place by the Hartree energy that scales like ${N\,\mathrm{log}N}$. The main purpose of this work is to introduce two disjunct semiclassical DPFT approximations whose computational costs for electronic structure calculations scale like $N$ and ${N\,\mathrm{log}N}$, respectively.

\subsection{\label{secGeneralformalism}General formalism of density--potential functional theory}

Orbital-free density functional theory (OF--DFT) seeks the extremum of the constrained density functional of the total energy 
\begin{align}\label{gsEnergy2}
E=E[n,\mu]=E_{\mathrm{kin}}[n]+E_{\mathrm{ext}}[n]+E_{\mathrm{int}}[n]
+\mu\left(N-\int(\d\vec r)\,n(\vec r)\right)\,,
\end{align}
which sums the kinetic energy, the energy due to the single-particle external potential, and the interaction energy for a quantum system of $N$ particles, associated with the chemical potential $\mu$. DPFT introduces the effective potential energy
\begin{align}\label{deltaEkin}
V(\vec r)=\mu-\frac{\delta E_{\mathrm{kin}}[n]}{\delta n(\vec r)}\,,
\end{align}
such that the Legendre transform
\begin{align}\label{LegendreTF}
  E_1[V-\mu]=E_{\mathrm{kin}}[n]
             +\int(\d\vec r)\,\big(V(\vec r)-\mu\big)\,n(\vec r)
\end{align}
of the kinetic energy functional $E_{\mathrm{kin}}[n]$ yields 
\begin{align}\label{EnergyVnmu}
E=E[V,n,\mu]=E_1[V-\mu]-\int(\d\vec r)\,n(\vec r)\,\big(V(\vec r)-V_{\mathrm{ext}}(\vec r)\big) +E_{\mathrm{int}}[n]+\mu N\,.
\end{align}
The $V$- and $n$-variations at the stationary point of $E[V,n,\mu]$ obey
\begin{align}
  \delta V: & \qquad \,n[V-\mu](\vec r)=
              \frac{\delta E_1[V-\mu]}{\delta V(\vec r)}\label{n}
\intertext{and}
 \delta n: & \qquad\label{V}
          V[n](\vec r)=V_{\mathrm{ext}}(\vec r)
          +\frac{\delta E_{\mathrm{int}}[n]}{\delta n(\vec r)}\,,
\end{align}
respectively. The $\mu$-variation, combined with Eq.~(\ref{n}), reproduces the particle-number constraint  
\begin{align}\label{PartNumConstraint}
\int(\d\vec r)\,n(\vec r)=N\,.
\end{align}
Equation~(\ref{n}) yields the particle density for any given $\mu$ and effective potential, e.g., the noninteracting density in the noninteracting case (${V=V_{\mathrm{ext}}}$). We find candidates of the ground-state density from the self-consistent solution of Eqs.~(\ref{n})--(\ref{PartNumConstraint}) for a given interaction functional $E_{\mathrm{int}}[n]$. The fundamental variables of DPFT are $V$ and $n$, in contrast both to the standard Hohenberg-Kohn DFT and to potential functional theory \cite{Peng2012,Cangi2013}, although the optimized effective potential approach to OF-DFT is related in spirit to DPFT, see Ref.~\cite{Peng2012} and references therein. In the following, we reiterate only the most important features and formulae of DPFT pertinent to the present work; details on DPFT can be found in Refs.~\cite{Englert1988,Englert1992,Trappe2016,Trappe2017,Trappe2023DFMPS,Trappe2021b}.

The main advantage of orbital-free DPFT is to avoid the not explicitly known density functional $E_{\mathrm{kin}}[n]$ of the kinetic energy in favor of its Legendre transform
\begin{align}
E_1[V-\mu]=\tr\{\mathcal E_T(H-\mu)\}\,.\label{tracef}
\end{align}
Here, ${H=H(\vec R,\vec P)}$ is the Hamiltonian of a system (at finite or zero temperature $T$), for which we may neglect couplings between the position operator $\vec R$ and the momentum operator $\vec P$, for example, of spin-orbit type \cite{Trappe2017,Englert1992,Trappe2021b}. The single-particle trace in Eq.~(\ref{tracef}) includes a degeneracy factor $g$, which can, for example, encode spin-multiplicity. We will use the explicitly known noninteracting version
\begin{align}\label{fHmuT}
  \mathcal E_T^{(0)}(A=H-\mu)
  =(-k_{\mathrm{B}}T)\,\ln{\left(1+\mathrm{e}^{-A/k_{\mathrm{B}}T}\right)}
\end{align}
of $\mathcal E_T$, which has a strong track record also for interacting systems \cite{Englert1988,Englert1992,Trappe2016,Trappe2017,Chau2018,Trappe2023DFMPS,Trappe2021b}, akin to neglecting the interacting part of the kinetic energy in constructions of the KS exchange--correlation functionals. Here and in the following we omit arguments of functions for brevity whenever expedient.

Equations~(\ref{tracef}) and (\ref{fHmuT}) permit explicit systematic approximations. In the following, we introduce two independent semiclassical schemes for DPFT. They can be benchmarked unambiguously for noninteracting systems or if the interaction functional is known or prescribed.

\subsection{\label{SectionST}Densities and energies from Suzuki--Trotter factorizations}

As derived in Refs.~\cite{Chau2018,Trappe2021b}, Eqs.~(\ref{n}) and (\ref{tracef}) at ${T=0}$ yield
\begin{align}\label{nSTA}
n(\vec r)=g\bok{\vec r}{\eta(\mu-H)}{\vec r}
         =g\Int\frac{\d t}{2\pi\I t}\,\e{\frac{\I t}{\hbar}\mu}\,
           \bok{\vec r}{U(t)}{\vec r}\,.
\end{align}
Here, we make use of the Fourier transform of the step function ${\eta(\ )}$, and the integration path from ${t=-\infty}$ to ${t=\infty}$ crosses the imaginary $t$ axis in the lower half-plane. Application-specific versions of Eq.~(\ref{nSTA}) can be derived. For example, linear dispersion (instead of the single-particle kinetic-energy operator ${T=\vec P^2/(2m)}$) in Ref.~\cite{Trappe2019} accounts for the Dirac cone in graphene, and momental densities are immediately accessible through the momentum-space version ${n(\vec p)=g\bok{\vec p}{\eta(\mu-H)}{\vec p}}$ of Eq.~(\ref{nSTA}). In any case, the time-evolution operator ${U(t)=\e{-\frac{\I t}{\hbar}H}}$ with Hamiltonian ${H=T+V}$ in Eq.~(\ref{nSTA}) can be systematically approximated by tailored Suzuki--Trotter factorizations. The crudest approximation of that sort is ${U\approx\e{-\frac{\I t}{\hbar}T}\e{-\frac{\I t}{\hbar}V}}$, which yields the TF density
\begin{align}\label{nTF}
n_{\mathrm{TF}}(\vec r)=\frac{g\,\Omega_D}{D\,(2\pi\hbar)^D}\big[2m\,(\mu-V(\vec r))\big]_+^{D/2}=\frac{g\,\Omega_D}{D\,(2\pi{\mathcal U}^2)^D}\big[2\,(\mu-V(\vec r))\big]_+^{D/2}\,,
\end{align}
with ${[z]_+=z\,\eta(z)}$ and the solid angle $\Omega_D$ in $D$ dimensions. Equation~(\ref{nTF}) exposes the units of energy ($\mathcal E$) and length ($\mathcal L$) via the dimensionless (mass-dependent) constant ${{\mathcal U}=\hbar^2/(m\,\mathcal L^2\,\mathcal E)}$. In all formulae of this work that exhibit ${\mathcal U}$, the quantities of energy are given in units of $\mathcal E$ and those of length in units of $\mathcal L$. For example, $\mu$ in Eq.~(\ref{nTF}) is implicit for $\mu/\mathcal E$, and $n_{\mathrm{TF}}(\vec r)$ comes in units of $\mathcal L^{-D}$. For the concrete examples in the sections below we use units of ${\mathcal E=\mbox{eV}}$ and ${\mathcal L=\mbox{\AA}}$ and choose the electron mass for $m$, such that ${\mathcal U\approx7.61996}$. Harmonic oscillator units, for instance, are implemented by ${\mathcal U=1}$.

In this work, we transform the approximation
\begin{align}\label{n3p}
n_{3'}(\vec r)=g\int(\d\vec a)\left(\frac{k_{3'}}{2\pi a}\right)^D J_D(2a\,k_{3'})
\end{align}
for the single-particle density into computationally more feasible expressions. Equation~(\ref{n3p}) is the quantum-corrected successor of $n_{\mathrm{TF}}$ \cite{Chau2018,Trappe2021b}, with the Bessel function $J_D(\,)$ of order $D$ and the effective Fermi wave number ${k_{3'}=\frac{1}{\hbar}\big[2m\big(\mu-V(\vec r+\vec a)\big)\big]_+^{1/2}}$. In contrast to the local TF density, whose computational cost scales with size $G$ of the numerical grid, $n_{3'}(\vec r)$ is a fully nonlocal expression, which samples the effective potential $V$ in a neighborhood of the position $\vec r$, such that its computational cost scales like~$G^2$. The accuracy of $n_{3'}$ and its associated kinetic energy
\begin{align}\label{Ekin3p}
E_{\mathrm{kin}}^{(3')}=\frac{g\,\Omega_D}{(2\pi{\mathcal U}^2)^D\,(2D+4)}\int(\d\vec r)\,\big[2\big(\mu-V(\vec r)\big)\big]_+^{\frac{D+2}{2}}
\end{align}
are sufficient for qualitative modeling of some basic chemistry applications like bond making and breaking. While such a capacity for small molecules is a clear improvement over the TF approximation, $n_{3'}$~should become quantitatively competitive with KS only for larger particle numbers.

Equation~(\ref{n3p}) is efficient enough for the calculation of isotropic densities. For anisotropic densities at zero temperature, we provide 
\begin{align}\label{n3pF}
n_{3'}^{\mathcal F}(\vec r)=\frac{g\,\Omega_D}{D\,(2\pi)^{D}}\mathcal F^{-1}\left\{\int(\d\vec r')\,\e{-\I\vec k\vec r'}\,\left[\frac{2}{{\mathcal U}}\big(\mu-V(\vec r')\big)-\frac{\vec k^2}{4}\right]_+^{D/2}\right\}(\vec r)\,.
\end{align}
in Appendix~\ref{AppendixAiryST} (see also Ref.~\cite{Trappe2021b}), which expresses $n_{3'}$ in terms of Fourier transforms $\mathcal F\{\,\}$ and is more efficient than Eq.~(\ref{n3p}), although the computational cost of $n_{3'}^{\mathcal F}$ still scales like $G^2$. In Appendix~\ref{AppendixAiryST} we also derive the finite-temperature version
\begin{align}\label{n3pT}
n_{3'}^T(\vec r)=\frac{g}{\Gamma[D/2]}\left(\frac{k_{\mathrm{B}}T}{2\pi{\mathcal U}}\right)^{D/2}\int_0^\infty\d y\,\mathcal{F}^{-1}\left\{\mathcal{F}\left\{f_y(\vec r')\right\}(\vec k)\,g_y^D(k)\right\}(\vec r)\,,
\end{align}
whose cost scales like ${G\,\log G}$ thanks to fast Fourier transforms, see Appendix~\ref{AppendixAiryST}, albeit with a (potentially) large prefactor that tends to increase with decreasing temperature. Here, $\Gamma(\,)$ denotes the Gamma function,
\begin{align}\label{fy}
f_y(\vec r')=\exp\left\{\big(\mu-V(\vec r')\big)/(k_{\mathrm{B}}T)-y\,\exp\left[\big(\mu-V(\vec r')\big)/(k_{\mathrm{B}}T)\right]\right\}\,,
\end{align}
and
\begin{align}\label{gyD}
g_y^D(k)=\int_0^\infty\d x\, x^{D/2-1}\,\exp\left[-y\,\exp\left(x+\kappa\right)\right]
\end{align}
is easily tabulated for the all required values of ${\kappa=(\hbar k)^2/(8m\,k_{\mathrm{B}}T)={\mathcal U}\,k^2/(k_{\mathrm{B}}T)}$, where $k$ is the magnitude of the wave vector $\vec k$ in Fourier space. With small enough temperature, $n_{3'}^T$ can be used in lieu of the ground-state density $n_{3'}$.

\subsection{\label{SectionAiry}Airy-averaged densities and energies}

From Refs.~\cite{Trappe2016,Trappe2017,Trappe2023DFMPS}, we recapitulate the most important expressions of our second approximation scheme, which derives from representing the trace in Eq.~(\ref{tracef}) by the classical phase space integral
\begin{align}\label{trH}
  \tr\{\mathcal E_T\big(A(\vec R,\vec P)\big)\}
  =g\int\frac{(\d\vec r)(\d\vec p)}{(2\pi\hbar)^D}\,
  \big[\mathcal E_T(A)\big]\W(\vec r,\vec p)\,.
\end{align}
The momentum integral over the Wigner function $\big[\mathcal E_T(A)\big]\W$ of $\mathcal E_T(A)$ can be approximated
by
\begin{align}
  \int(\d\vec p)\,\big[\mathcal E_T( A)\big]\W(\vec r,\vec p)
  \cong\int(\d\vec p)\,{\left\langle\!\mathcal E_T\big(\tilde{A}\W\big)
  -\frac{\hbar^2(\nab^2V)}{12m}\mathcal E_T''
  \big(\tilde{A}\W\big)\!\right\rangle}_{\hspace*{-0.2em}\mathrm{Ai}}\,.\label{trfA4}
\end{align}
Here, ${\tilde{A}\W(\vec r,\vec p)=H\W(\vec r,\vec p)-\mu-x\, a(\vec r)}$, with the Wigner function ${H\W(\vec r,\vec p)=\vec p^2/(2m)+V(\vec r)}$ of the single-particle Hamiltonian and ${a(\vec r)=|\hbar\nab V(\vec r)|^{2/3}/\big(2m^{1/3}\big)}$. We also call ${\mw{f}_\mathrm{Ai}=\int_{-\infty}^\infty\d x\,\mathrm{Ai}(x)f(x)}$, with the Airy function $\mathrm{Ai}(\ )$, the Airy average of the function $f$, and `$\cong$' stands for an approximation that reproduces the leading gradient correction exactly. Equation~(\ref{trfA4}) holds not only for $\mathcal E_T(A)$, but for any function of $A$ that has a Fourier transform. Equation (\ref{trfA4}) is exact up to the leading gradient correction $\big(\mathcal O(\nab^2)\big)$, and thus presents a systematic correction to the TF approximation, which is recovered in the uniform limit $\big(\mathcal O(\nab^0)\big)$. However, the `Airy-average' in Eq.~(\ref{trfA4}) also contains higher-order gradient corrections that enter through the Moyal products from the power
series expansion of $\mathcal E_T(A)$ in Eq.~(\ref{trH}). These higher-order corrections are responsible for the almost exact behavior of particle densities across the boundary between classically allowed and forbidden regions, where the TF approximation can fail epically (even if supplemented with the leading gradient correction \cite{Trappe2017}). Eventually, we obtain the Airy-averaged particle densities for one-, two-, and three-dimensional geometries by combining Eqs.~(\ref{n}), (\ref{tracef}), (\ref{fHmuT}), (\ref{trH}), and (\ref{trfA4}) and by evaluating the momentum integral of Eq.~(\ref{trfA4}). The 1D and 2D situations are covered extensively in Refs.~\cite{Trappe2023DFMPS} and \cite{Trappe2016,Trappe2017}, respectively. Here, we derive the explicit Airy-averaged expressions for energies and densities in 3D. 

The Airy-averaged 2D ground-state densities $n_{\mathrm{Ai}}^{T=0}$ of Ref.~\cite{Trappe2017} exhibit unphysical oscillations in the vicinity of positions~$\vec r$ where ${\nab V(\vec r)=0}$. By introducing a small but finite temperature $T$, we obtain densities that are well-behaved everywhere. The analogous derivation of the finite-temperature 3D expression
\begin{align}\label{nAiT}
n_{\mathrm{Ai}}^{T}(\vec r)=u_0\times\left\{
\begin{array}{ll}
\frac{a(\vec r)}{k_{\mathrm{B}}T}\int\d x\,\mathcal{A}(x)\,\left[u_1\,\mathrm{Li}_{1/2}\Big(-\mathrm{e}^{-\nu_x(\vec r)}\Big)+u_2(\vec r)\,\mathrm{Li}_{-3/2}\Big(-\mathrm{e}^{-\nu_x(\vec r)}\Big)\right] & ,\;a(\vec r)>0\\
u_1\,\mathrm{Li}_{1/2}\Big(-\mathrm{e}^{-\nu_0(\vec r)}\Big)+u_2(\vec r)\,\mathrm{Li}_{-3/2}\Big(-\mathrm{e}^{-\nu_0(\vec r)}\Big) & ,\;a(\vec r)=0
\end{array}
\right.
\end{align}
and details of its numerical implementation are provided in Appendix~\ref{AppendixAiryST}. We denote the polylogarithm of order $s$ by $\mathrm{Li}_{s}(\,)$ and the negative anti-derivative of the Airy function by ${\mathcal A(x)=\int_x^\infty\d y\,\mathrm{Ai}(y)}$. We set ${\mathcal A(x)=0}$ for ${x>100}$, use the asymptotic approximation ${\mathcal A(x)\approx 1-\cos\left(\pi/4+(2/3)\,|x|^{3/2}\right)/\left(\sqrt{\pi}\,|x|^{3/4}\right)}$ for ${x<-150}$, and tabulate $\mathcal A(x)$ for ${-150\le x\le100}$. We have
\begin{align}
u_0&=-\frac{g}{\big(2\pi\big)^{3/2}},\label{beta0}\\
u_1&=\left(\frac{k_{\mathrm{B}}T}{{\mathcal U}}\right)^{3/2},\\
u_2(\vec r)&=-\frac{\Delta V(\vec r)}{12}\left(\frac{k_{\mathrm{B}}T}{{\mathcal U}}\right)^{1/2}\,,
\end{align}
and 
\begin{align}
\nu_x(\vec r) = \frac{1}{k_{\mathrm{B}}T}\big(V(\vec r)-\mu-x\,a(\vec r)\big)\,.\label{nux}
\end{align}

In Appendix~\ref{AppendixAiryST} we also derive ${E_1^{\mathrm{Ai,T}}[V-\mu]}$ and the corresponding value
\begin{align}
E_{\mathrm{kin}}^{\mathrm{Ai},T}=E_1^{\mathrm{Ai},T}[V-\mu]-\int(\d\vec r)\,V(\vec r)\,n_{\mathrm{Ai}}^T(\vec r)+\mu N 
\end{align}
of the Airy-averaged kinetic energy at the stationary point of $E$. However, the ground-state kinetic energy
\begin{align}
E_{\mathrm{kin}}^{\mathrm{Ai}}&=\frac{g}{(2\pi\hbar)^3}\int(\d\vec r)\,{\left\langle\!4\pi\int\d p\,\frac{p^4}{2m}\left[f\big(\tilde{A}\W\big)-\frac{\hbar^2(\nab^2V)}{18m}f''\big(\tilde{A}\W\big)\right]\;\!\right\rangle}_{\hspace*{-0.2em}\mathrm{Ai}}\nn\\
&=\frac{g}{4\pi^2{\mathcal U}^2}\int(\d\vec r)\left\{
\begin{array}{ll}
\int_0^\infty\d x\,\mathrm{Ai}\left(x+\frac{V-\mu}{a}\right)\,\left[\frac{\sqrt{{\mathcal U}}}{5}\,(2\,a\,x)^{5/2}-\frac{{\mathcal U}^{3/2}\,(\nab^2V)}{6}\,(2\,a\,x)^{1/2} \right] & ,\;a(\vec r)>0\\
\left[\frac{\sqrt{{\mathcal U}}}{5}\,\left[2\,(\mu-V)\right]_+^{5/2}-\frac{{\mathcal U}^{3/2}\,(\nab^2V)}{6}\,\left[2\,(\mu-V)\right]_+^{1/2} \right] & ,\;a(\vec r)=0
\end{array}
\right.\label{EkinAi0}
\end{align}
is much better behaved numerically than $E_{\mathrm{kin}}^{\mathrm{Ai},T}$, such that we utilize $E_{\mathrm{kin}}^{\mathrm{Ai}}$ for calculating the DPFT energies also for systems at finite temperature as long as $T$ is small enough. We obtain Eq.~(\ref{EkinAi0}) after suitable integrations by part and evaluation of the momentum integral in
\begin{align}
E_{\mathrm{kin}}&=\tr\{T\,\eta(\mu-H)\}\nn\\
&=\frac{g}{(2\pi\hbar)^3}\int(\d\vec r)\int(\d\vec p)\,\frac{\vec p^2}{2m}\,\big[f( A)\big]_W(\vec r,\vec p)\nn\\
&\cong E_{\mathrm{kin}}^{\mathrm{Ai}}=\frac{g}{(2\pi\hbar)^3}\int(\d\vec r)\int(\d\vec p)\frac{\vec p^2}{2m}\int\d x\, \mathrm{Ai}(x)\left[f\big(\tilde{A}_W\big)-\frac{\hbar^2(\nab^2V)}{12m}\frac{D-1}{D}f''\big(\tilde{A}_W\big)\right]\,,\label{specialeAWp2}
\end{align}
reported in Ref.~\cite{Trappe2017}, for ${f(A)=\eta(-A)}$, ${f''(A)=\delta'(-A)}$ and ${D=3}$. For constant effective potential $V$ (then, ${a=0}$), Eq.~(\ref{EkinAi0}) recovers the TF kinetic energy, cf.~Eq.~(\ref{Ekin3p}). Like in the 2D case, $E_{\mathrm{kin}}^{\mathrm{Ai}}$ does not suffer from unphysical oscillations as ${a\to0}$. The computational costs of $n_{\mathrm{Ai}}^{T}$, $E_1^{\mathrm{Ai}}$, and $E_1^{\mathrm{Ai,T}}$ all scale with grid size $G$ (and thus linearly with particle number $N$ for electronic matter), just like the TF approximation, but in the current implementation the prefactor due to the Airy-average usually comes in at about $10^3$--$10^5$ for high-precision calculations.

\section{\label{secNuclearBackground}Nuclear background and interaction energies}

For the nuclear Coulomb potential, $n_{3'}$ diverges logarithmically, like $\log(1/r)$ as $r\to0$, a stark improvement upon the $r^{-3/2}$-scaling of $n_{\mathrm{TF}}$. This observation and the scaling behavior of $n_{\mathrm{Ai}}^T$ will be covered in detail elsewhere. Our numerical studies suggest that $n_{\mathrm{Ai}}^T$ diverges like $n_{\mathrm{TF}}$ at singularities of the (effective) potential, which is not surprising since $n_{\mathrm{Ai}}^T$ is a gradient expansion built on $n_{\mathrm{TF}}$ as the leading term. When employing the semiclassical densities introduced here, we therefore have to replace the nuclei's Coulomb potentials by pseudopotentials. Alternative DFT formulations that can cope with unregularized singular potentials, for example, via proper incorporation of the Scott correction \cite{Englert1988,Buchwald1989} or suitable basis function expansions are currently being developed \cite{Englert2023DFMPS,Trappe2023}. For all-electron calculations we replace the bare Coulomb potential of a nucleus of charge $Z$ by the smooth function
\begin{align}\label{aePP}
\mbox{\AE}_\alpha^{(Z)}(r)=-Z\,W\,\left((0.923 + 1.568\,\alpha)\,\exp\left(-(0.241 + 1.405\,\alpha)^2\,r^2\right) + \frac{\mathrm{Erf}(\alpha\, r)}{r}\right)\,,
\end{align}
which recovers the Coulomb potential for ${\alpha\to\infty}$, see Ref.~\cite{GonzalezEspinoza2016}. Of course, other replacements of the ionic Coulomb potential are possible \cite{Gygi2023}, but Eq.~(\ref{aePP}) suffices for the proof-of-principle calculations in this work. We have ${W=\mathcal U\,\mbox{{\AA}}/a_0\approx14.39965}$ dimensionless upon expressing all quantities of length in {\AA} and all energies in eV. $\mathrm{Erf}(\,)$ denotes the error function. By demanding ${\mbox{\AE}_\alpha^{(Z)}(0)=-2\,Z\,W/\Delta x}$, where $\Delta x$ is the lattice constant of the numerical grid, we determine an appropriate value of $\alpha$. This condition is implied by the most simple regularization
\begin{align}\label{CPP}
\mathcal{C}^{(Z)}(r)=-\frac{Z\,W}{\mathrm{max}\{r,\Delta x/2\}}
\end{align}
of the Coulomb potential, which coincides with ${-Z\,W/r}$ (except at the origin) and is smooth at ${r=\Delta x}$. For calculating valence densities with two (three) electrons per Mg (Al) atom we enlist the GGA pseudopotentials that accompany the OF--DFT package PROFESS \cite{Chen2015,Dieterich2017b}. In all our calculations we treat all electrons as unpolarized (also systems with an odd number of electrons); hence,~$g=2$.

Approximations of the interaction functionals and their derivatives in Eq.~(\ref{V}) can, in principle, be obtained consistently within the same approximation schemes that yield the semiclassical approximations of Eq.~(\ref{n}). This agenda is beyond the scope of this article, but the structural similarity between DPFT and KS--DFT invites the use of established KS exchange--correlation functionals for calculating Coulomb-interacting systems in 3D with DPFT. For the DPFT calculations in this work we used the LDA \cite{Vosko1980} and PBE \cite{Perdew1996,Perdew1997Erratum} implementations from the LIBXC library \cite{Lehtola2018}.

Kohn--Sham DFT calculations were performed in Gaussian~16 \cite{Frisch2016} and in Abinit \cite{Gonze2020}. We used the 6-31g(d,p) basis set for atoms and dimers and D95 for the 201-atom Al nanoparticle. Gaussian were done with LSDA \cite{Slater1974,Vosko1980} and PBE functionals. All Abinit calculations were performed with PBE, a plane wave cutoff of $500\,$eV, and local pseudopotentials from \cite{Huang2008,Legrain2015}. In Abinit, the systems were placed in a large vacuum box (e.g., $36\,\mbox{\AA}\vphantom{A}^3$ for the nanoparticle), and the calculations done at the Gamma point. Coupled cluster calculations for atoms were performed in Gaussian~16 with singles, doubles, and perturbative triples $\big(\mbox{CCSD(T)}\big)$, using the aug-cc-pv5z basis set.

\section{\label{Results}Results for atoms, dimers, and nanoparticles}

We first establish the quality of the DPFT densities $n_{3'}^{T\ge0}$ and $n_{\mathrm{Ai}}^{T>0}$ for single atoms and dimers by benchmarking against KS and coupled-cluster results. We calculate the valence densities and the all-electron densities as declared in Sec.~\ref{secNuclearBackground}; all results shown are for the valence density and are obtained with PBE, unless explicitly stated otherwise; all-electron densities are labeled by `(\AE)' throughout this work; the employed numerical integration grids are declared in Table~\ref{IntegrationGrids} in Appendix~\ref{NumericalDetails}.

\subsection{\label{secAtoms}Single Atoms}

\begin{figure}[htb!]
\begin{center}
\includegraphics[height=0.4\linewidth]{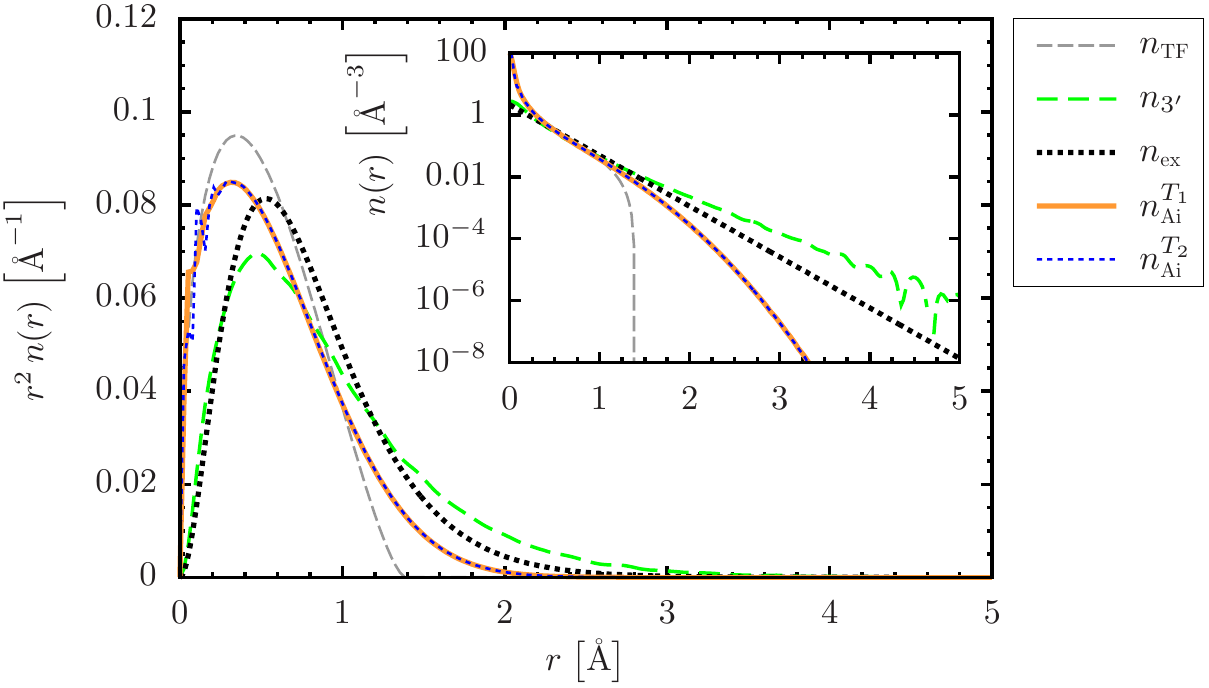}
\caption{\label{H}Comparison of DPFT electron densities with the exact density $n_{\mathrm{ex}}$ of the hydrogen atom (main plot, rescaled abscissa). We show the quasi-classical TF density and its quantum-corrected successors, the semiclassical densities $n_{3'}$ as well as $n_{\mathrm{Ai}}^T$ at ${k_{\mathrm B}T_1=0.1\,\mathrm{eV}}$ and ${k_{\mathrm B}T_2=10^{-6}\,\mathrm{eV}}$, respectively. The semiclassical densities reach deep into the classically forbidden region, where $n_{3'}$ approximates the characteristic exponential decay (inset).}
\end{center}
\end{figure}

First, we consider the hydrogen atom that hosts a single electron, for which the explicit noninteracting kinetic energy functional in Eq.~(\ref{fHmuT}) is exact. We can therefore unambiguously benchmark the approximate semiclassical DPFT densities in Eqs.~(\ref{n3p}), (\ref{n3pT}), and (\ref{nAiT})---albeit for ${N=1}$. In Fig.~\ref{H} we find a markedly improved density tail of $n_{3'}$ compared with $n_{\mathrm{TF}}$, deep into the classically forbidden region of the Coulomb potential, for which we employ Eq.~(\ref{aePP}). There is no difference (to the eye) when using the pseudopotential $\mathcal{C}^{(1)}$ of Eq.~(\ref{CPP}) instead. Deviations from the pure exponential decay of the exact density $n_{\mathrm{ex}}$ are expected since the semiclassical approximation $n_{3'}$ generally performs better for larger~$N$. The same holds for $n_{\mathrm{Ai}}^T$, where we find ${k_{\mathrm B}T_1=0.1\,\mathrm{eV}}$ sufficiently small for targeting the ground-state density, as we judge from comparing with the density at ${k_{\mathrm B}T_2=10^{-6}\,\mathrm{eV}}$. As a general strategy for selecting low enough temperatures that yield density profiles close to the ground-state density, we start with high temperatures that incur small computational cost and decrease $T$ until the change in density is negligible; for different systems this happens at different temperatures. As expected the scaling behavior of $n_{\mathrm{Ai}}^T$ near the singularity of the Coulomb potential is similar to that of $n_{\mathrm{TF}}$.

\begin{figure}[htb!]
\begin{center}
\includegraphics[height=0.4\linewidth]{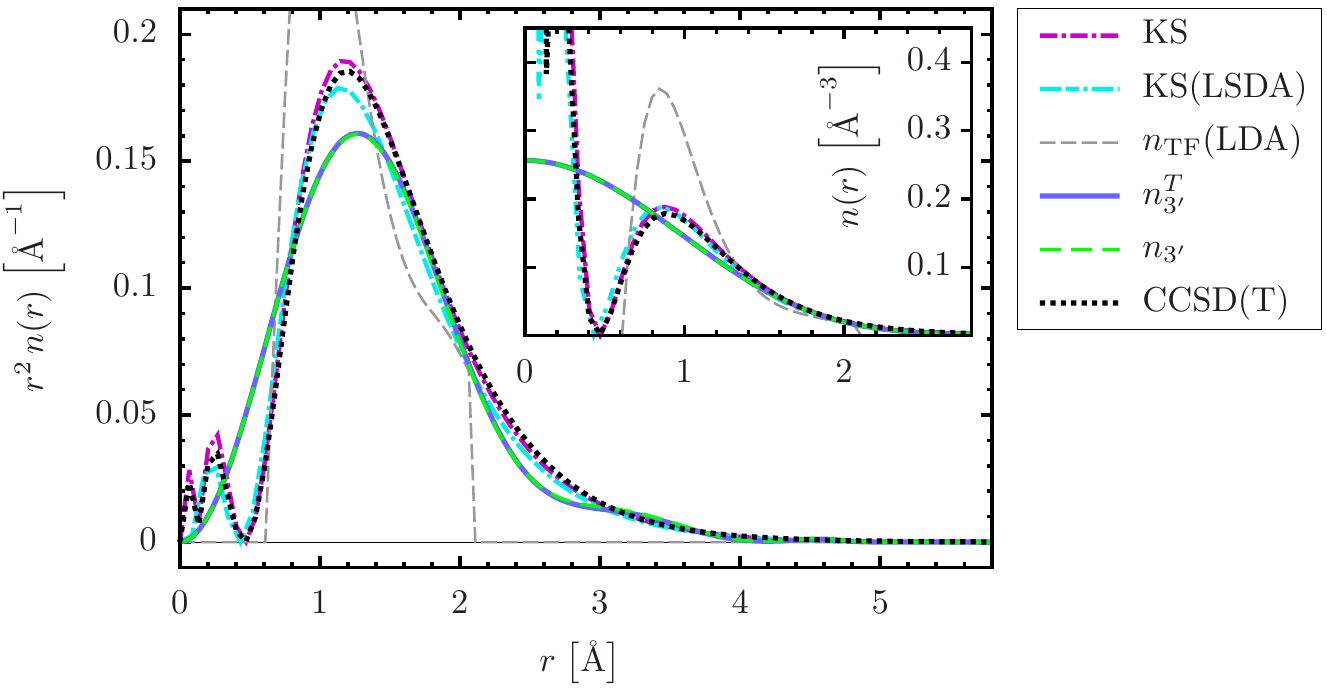}
\caption{\label{Al}DPFT electron densities like in Fig.~\ref{H}, but for the valence electrons of aluminum (main plot, rescaled abscissa). The larger deviations of $n_{3'}$ near the nucleus (inset) can generally be considered benign when calculating electronic properties of atomic matter, for which the valence region is key.}
\end{center}
\end{figure}

While any semiclassical method can be expected to fail for ${N=1}$ particle, we find in Fig.~\ref{Al} that $n_{3'}$ and $n_{3'}^T$ capture the ${(N=3)}$-electron valence density of an aluminum atom adequately---especially in the regions of the atom that are important for determining bond properties. Both the KS densities with (i) LSDA and (ii) PBE exchange--correlation functional are reasonably close to the quasi-exact result from a coupled-cluster calculation $\big(\mbox{CCSD(T)}\big)$. The discontinuous derivative of $n_{\mathrm{TF}}$ makes convergence of the self-consistent DPFT loop troublesome when using PBE (and we thus employ $n_{\mathrm{TF}}$ with LDA), but converging $n_{3'}$ with PBE is unproblematic. A temperature~$T$ corresponding to ${k_{\mathrm{B}}T=0.1\,}$eV brings $n_{3'}^T$ close enough to its ground-state version $n_{3'}$. We find $n_{3'}$ to (i) give an approximate average account of the quantum oscillations near the nucleus and (ii) approach the quality of the KS(LSDA) results in the valence region, with $n_{3'}$ being superior to the KS density in some parts of the evanescent region and inferior in others. This is no small feat when bearing in mind that $n_{3'}$ is only the first step in a hierarchy of systematic improvements upon the TF approximation.

\FloatBarrier

\begin{figure}[htb!]
\begin{center}
\includegraphics[height=0.4\linewidth]{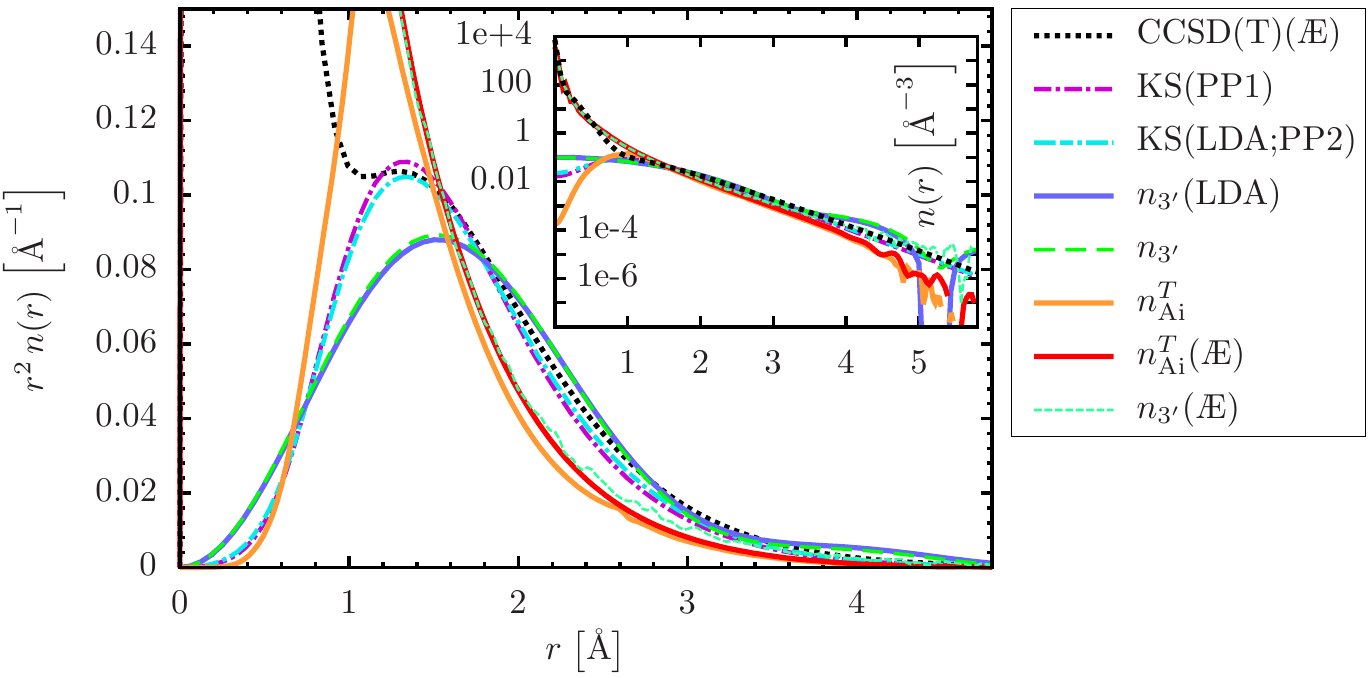}
\caption{\label{Mg}DPFT electron densities like in Fig.~\ref{Al}, but for magnesium (main plot, rescaled abscissa). The quasi-exact density in the bonding region is reasonably well matched over several orders of magnitude (inset). The KS calculations are rather insensitive to different pseudopotentials: `PP1' labels the density obtained from the GGA pseudopotential of Ref.~\cite{Huang2008} (deployed together with PBE) and closely aligns with the result that stems from an alternative pseudopotential (labeled `PP2', deployed together with LDA) from Ref.~\cite{Legrain2015}. We used ${k_{\mathrm B}T=0.05\,\mathrm{eV}}$ for both the valence density $n_{\mathrm{Ai}}^{T}$ and the all-electron density $n_{\mathrm{Ai}}^{T}(\mbox{\AE})$.}
\end{center}
\end{figure}

Figure~\ref{Mg} illustrates valence densities as well as all-electron densities of a magnesium atom. Both of our semiclassical approaches capture the quasi-exact CCSD(T) all-electron density over about eight orders of magnitude, and in particular the exponential decay in the valence region, where the valence density $n_{\mathrm{Ai}}^{T}$ follows its all-electron version and aligns with the all-electron density $n_{3'}(\mbox{\AE})$. In the tail, both $n_{3'}$ and $n_{3'}(\mbox{\AE})$ oscillate closely around the KS and CCSD(T) densities, similar to what we observe for aluminum in Fig.~\ref{Al}. We observe similar profiles for $n_{3'}$ with LDA and PBE, respectively, except deep in the classically forbidden region, where $n_{3'}$ can become negative, such that its derivatives make the use of PBE unreliable. In summary, then, various reasonable settings (LDA, PBE, different pseudopotentials, valence- or all electron density, box sizes, and resolution) all produce a similar and coherent picture for single atoms, which invites us to move on to dimers.

\FloatBarrier

\subsection{\label{secDimers}Dimers}  

\begin{figure}[htb!]
\begin{center}
\includegraphics[height=0.4\linewidth]{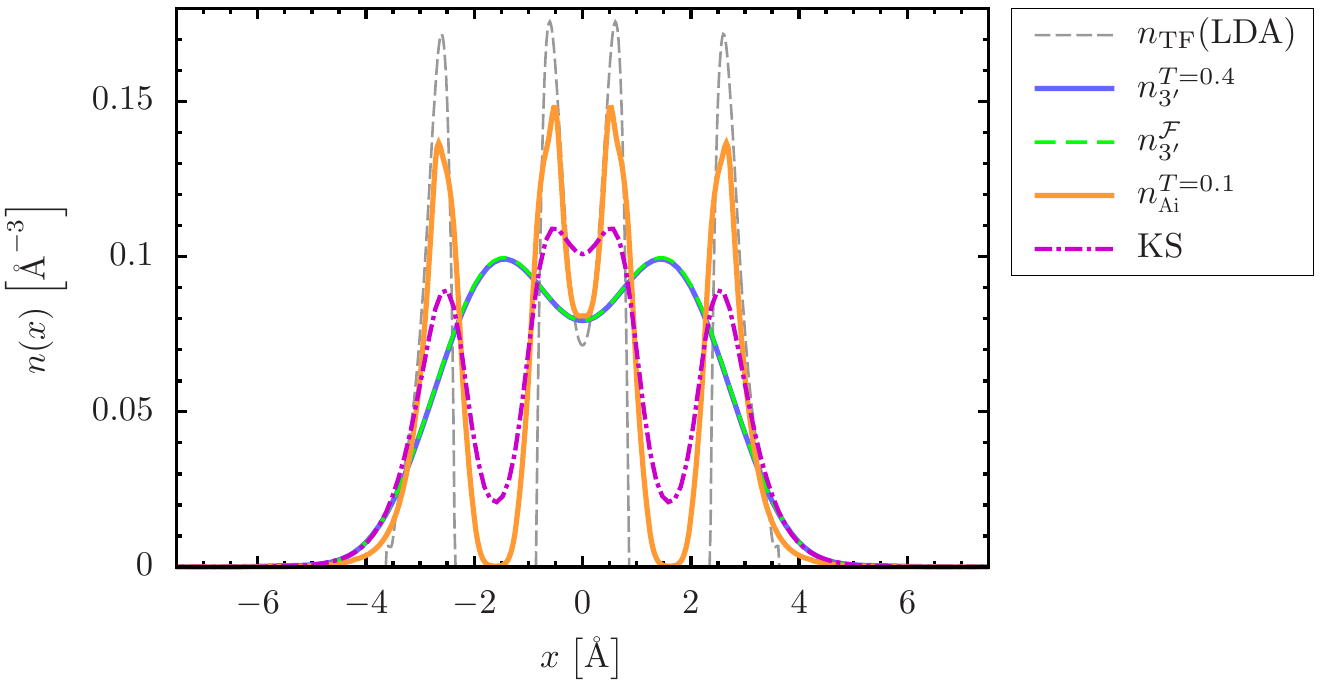}
\caption{\label{pseudoMg2} The valence densities from DPFT and KS calculations along the nuclear axis of a magnesium dimer with nuclei positioned at ${x\approx\pm1.60\,\mbox{\AA}}$, i.e., separated by the approximate equilibrium distance of $3.21\,${\AA}. We show the densities $n_{3'}^T$ for a temperature ${T=0.4\,\mbox{eV}/k_{\mathrm{B}}}$---low enough for matching $n_{3'}^{\mathcal F}$ almost exactly---and $n_{\mathrm{Ai}}^T$ for ${k_{\mathrm{B}}T=0.1\,}$eV. For anisotropic high-resolution setups like our Mg-dimer simulation here, $n_{3'}^T$ is computationally more efficient than $n_{3'}$ and $n_{3'}^{\mathcal F}$. Figure~\ref{ContourPlots} in Appendix~\ref{NumericalDetails} displays contour plots of $n_{3'}^T$ and $n_{\mathrm{Ai}}^T$ for the Mg-dimer.}
\end{center}
\end{figure}

Next, we benchmark dimer densities
that include the effects of bonding. We begin with the ground-state valence density of a magnesium dimer with nuclei separated by the approximate equilibrium distance. This case reveals several properties of our semiclassical density formulae, which, for ${N=4}$, do not yet operate in a truly semiclassical regime. Significant differences to the KS density can therefore be expected. Indeed, Fig.~\ref{pseudoMg2} shows that $n_{3'}^T$ produces an approximate average account of the characteristic quantum-mechanical oscillations of the KS electron density. As a side-effect, $n_{3'}^T$ conjures up maxima at the nuclei instead of minima, consistent with what we report for a single Al atom in the inset of Fig.~\ref{Al}. This is usually unproblematic since the valence region is of primary interest in most applications. Furthermore, like for single atoms, the density tails of $n_{3'}^T$ closely align with the KS predictions. To a lesser extent this also holds for $n_{\mathrm{Ai}}^T$, which, on the other hand, proves superior to $n_{3'}$ in capturing the global pattern of the density distribution---heralded by the TF density, which is the leading term of the gradient-corrected $n_{\mathrm{Ai}}^T$. The stark differences between $n_{3'}^T$ and $n_{\mathrm{Ai}}^T$ highlight the dissimilarity of their semiclassical origins.

\begin{figure}[htb!]
\begin{center}
\includegraphics[height=0.4\linewidth]{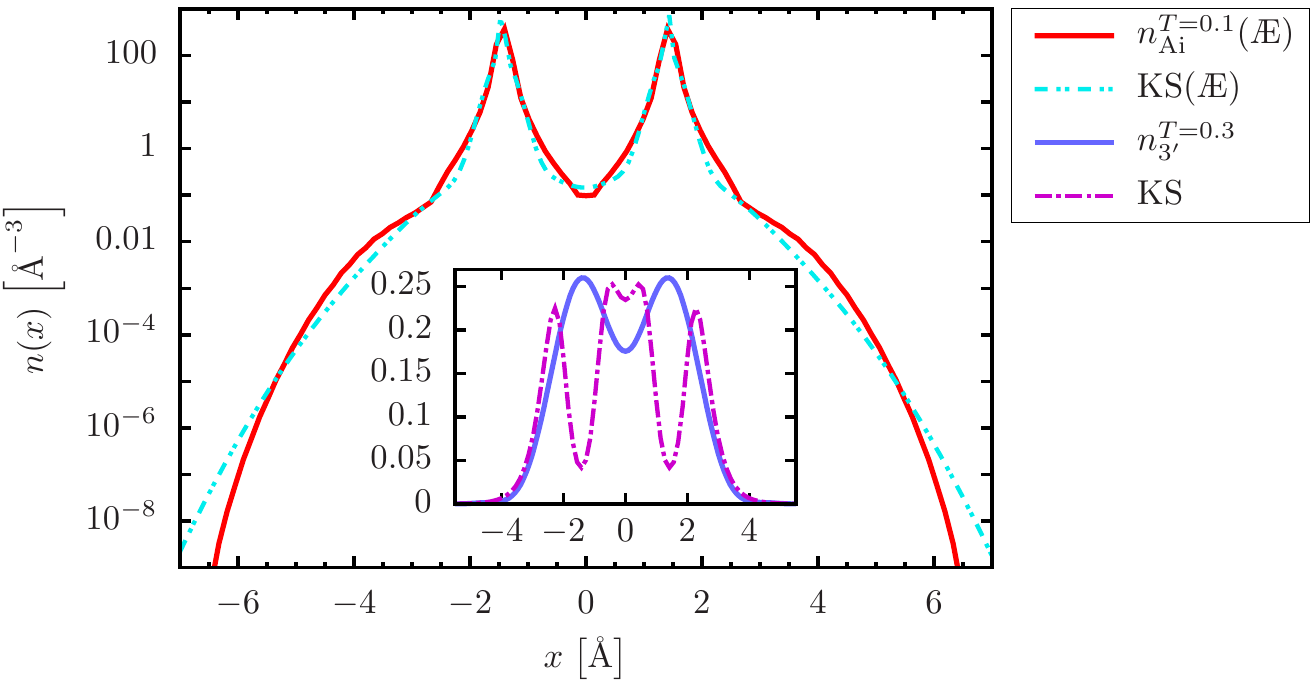}
\caption{\label{Al2} All-electron densities (main plot) and valence densities (inset) of the aluminum dimer with nuclei separated by the approximate equilibrium distance of $2.86\,${\AA}.}
\end{center}
\end{figure}

The valence-density profile from $n_{3'}^T$ for the Al-dimer in Fig.~\ref{Al2} (inset) follows a pattern similar to that of the Mg-dimer in Fig.~\ref{pseudoMg2}, but better performance can be expected from our semiclassical densities for larger particle numbers. Indeed, as we show in Fig.~\ref{Al2} (main plot), it is reassuring that the semiclassical expression $n_{\mathrm{Ai}}^T$ delivers a reasonable approximation of the KS density of all 26 electrons of the Al-dimer across more than eight orders of magnitude.

\FloatBarrier

\subsection{\label{secNanoparticles}Aluminum nanoparticle}

\begin{figure}[htb!]
\begin{minipage}{0.72\linewidth}
\includegraphics[width=\linewidth]{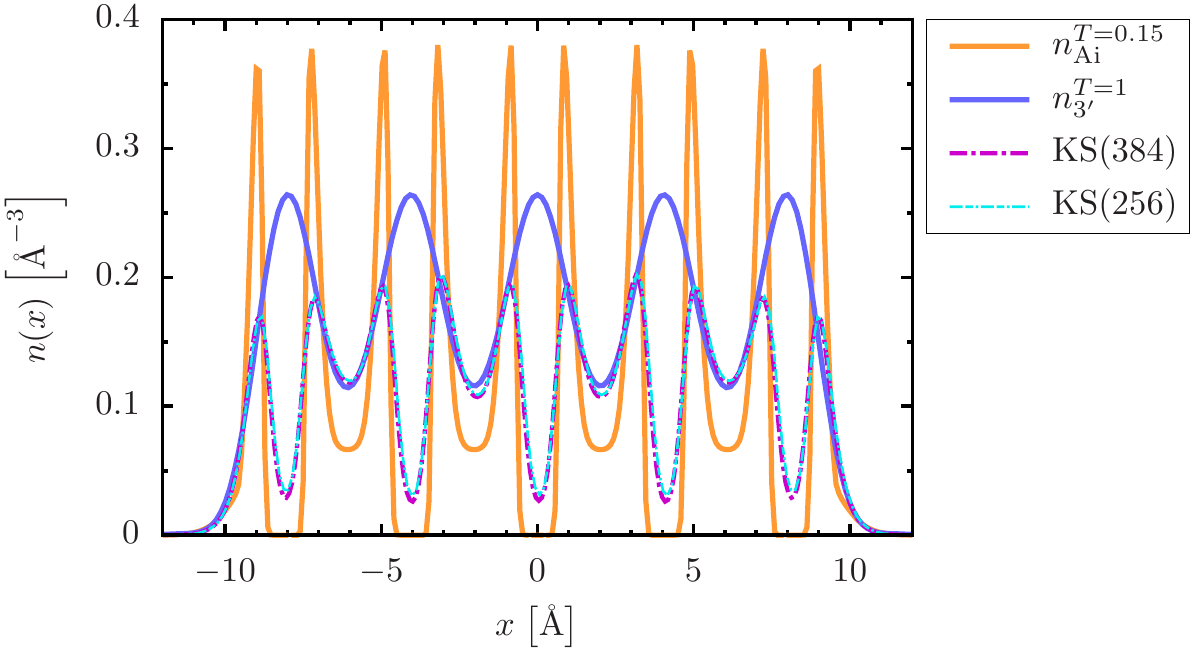}
\end{minipage}
\begin{minipage}{0.275\linewidth}
\includegraphics[width=\linewidth]{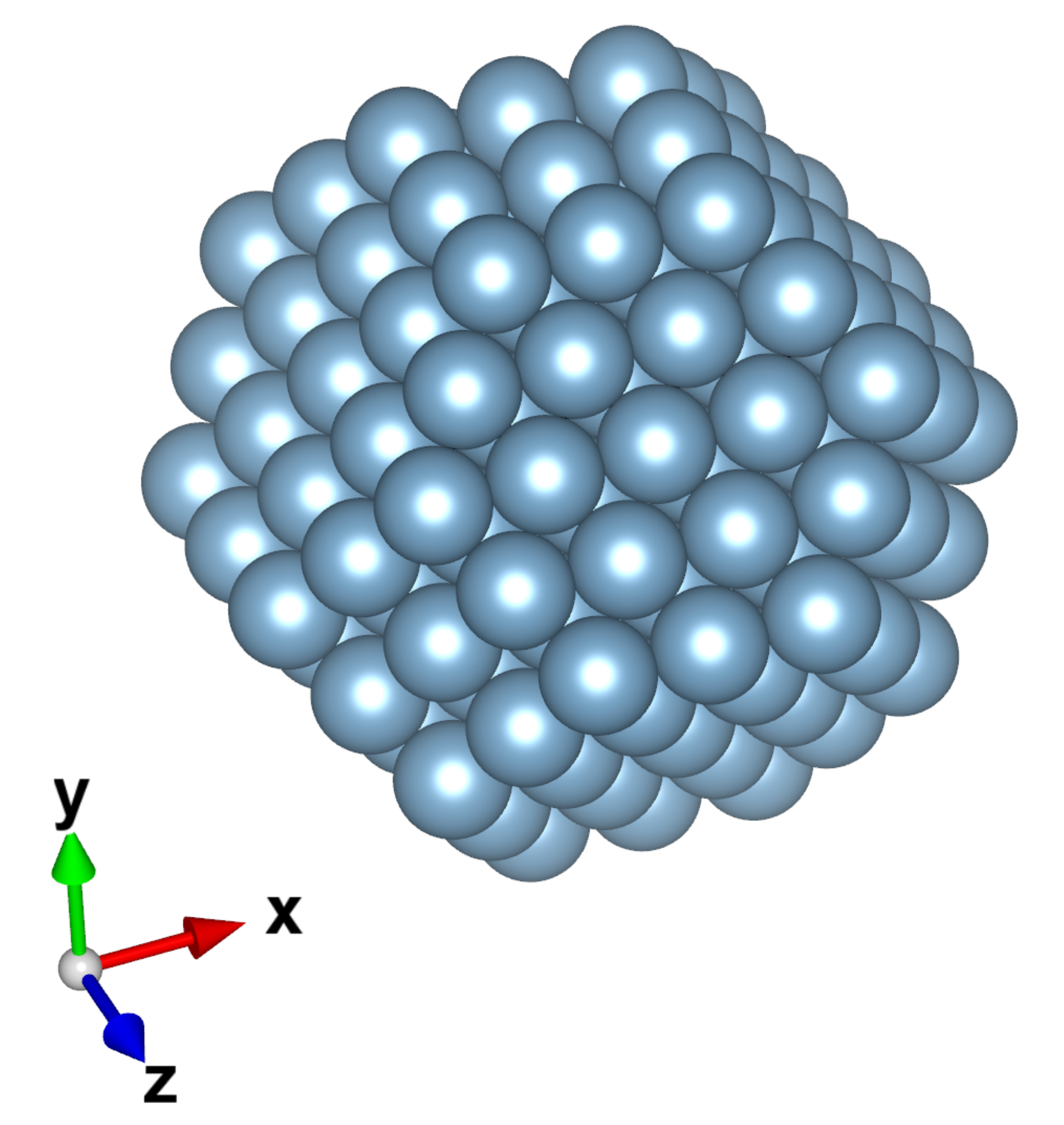}
\end{minipage}
\caption{\label{201Aln3pT}Cut through the valence densities along the x-axis (left) of a Wulff-shaped 201-Al-atom nanoparticle (right), as predicted by the semiclassical DPFT formulae $n_{\mathrm{Ai}}^T$ and $n_{3'}^{T}$. The KS densities were computed on grids of $256^3$ and $384^3$ points, respectively. The discrepancies between $n_{3'}^{T}$ and the KS densities in the bond region are less pronounced if electrons of several atoms contribute to the electron density between any two nuclei---rather than only two atoms like for the dimers in Figs.~\ref{pseudoMg2} and~\ref{Al2}.}
\end{figure}

\begin{figure}[htb!]
\includegraphics[height=0.4\linewidth]{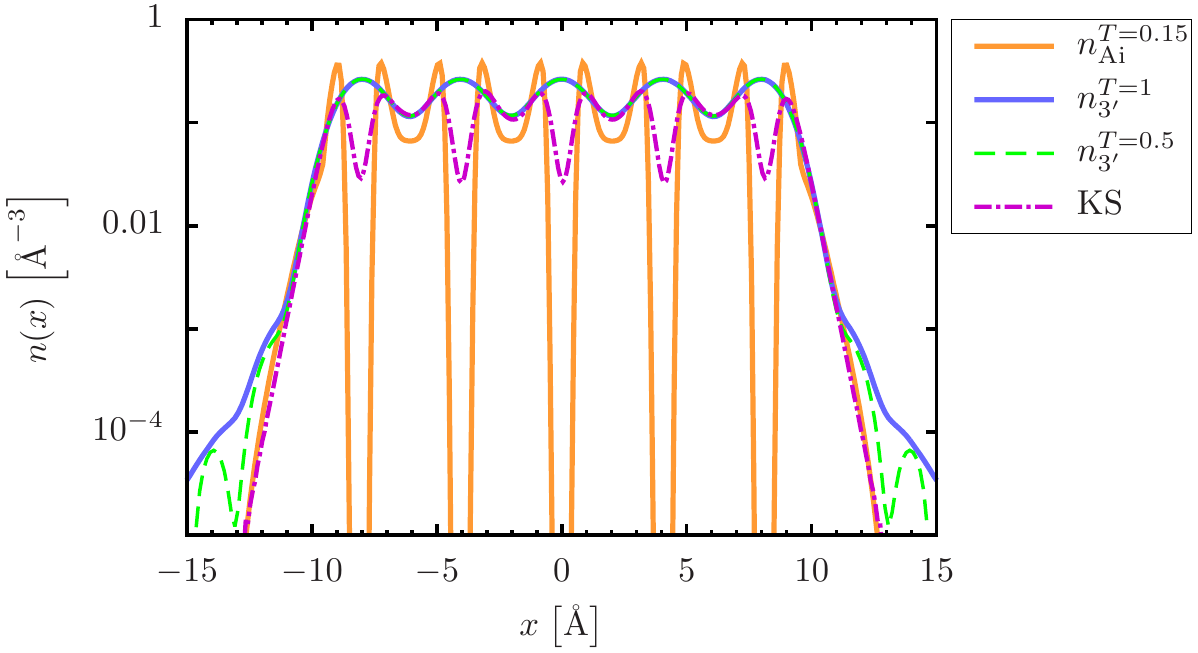}
\caption{\label{201Aln3pTlogscale}The valence densities of $n_{\mathrm{Ai}}^{T=0.15}$ and $n_{3'}^{T=1}$ like in Fig.~\ref{201Aln3pT}, but on a logarithmic scale. Only the transitional region toward vacuum exhibits significant deviations between $n_{3'}^{T}$ at ${k_{\mathrm{B}}T=1\,\mathrm{eV}}$ and $n_{3'}^{T=0.5}$, respectively. There, $n_{3'}$ generally becomes unreliable anyway, while the performance of $n_{\mathrm{Ai}}^{T=0.15}$ incidentally leaves nothing to be desired.}
\end{figure}

Finally, we benchmark $n_{\mathrm{Ai}}^T$ and $n_{3'}^{T}$ against KS densities of a Wulff-shaped nanoparticle composed of 201 aluminum atoms, whose nearest-neighbor distance is fixed to the approximate bulk equilibrium separation of $2.86\,${\AA} of the fcc structure. The particle is shown in Fig.~\ref{201Aln3pT}. Figures~\ref{201Aln3pT} and \ref{201Aln3pTlogscale} show the valence densities of the nanoparticle, analogous to the dimer densities in Fig.~\ref{pseudoMg2} and the inset of Fig.~\ref{Al2}. Overall, the patterns of the DPFT densities are what we naively expect based on the dimer simulations in Sec.~\ref{secDimers}: $n_{3'}^{T}$ captures the internuclear and boundary regions, but exhibits unphysical oscillations into negative densities in the far tails and fails to decrease toward the nuclei. In line with our results for the Mg dimer, $n_{\mathrm{Ai}}^T$ predicts the overall density of the nanoparticle more accurately, especially deep into the evanescent region, but overestimates the amplitudes of the density modulations in the bulk. Interestingly, when comparing the bond regions in Figs.~\ref{Al2} and \ref{201Aln3pT}, we find $n_{3'}^{T}$ to perform much better for the nanoparticle than for the Al dimer. The improved behavior of $n_{3'}^{T}$ stems from its nonlocality, where information is drawn from an extended region around the focal point $\vec r$. Given up to six nearest-neighbor atoms (i.e., 21 valence electrons) in the vicinity of each bond region of the nanoparticle, it is then not surprising that the averaging effect of $n_{3'}^{T}$ improves the electron densities relative to the bond region of the aluminum dimer, where only six valence electrons contribute.

\begin{figure}[htb!]
\includegraphics[height=0.4\linewidth]{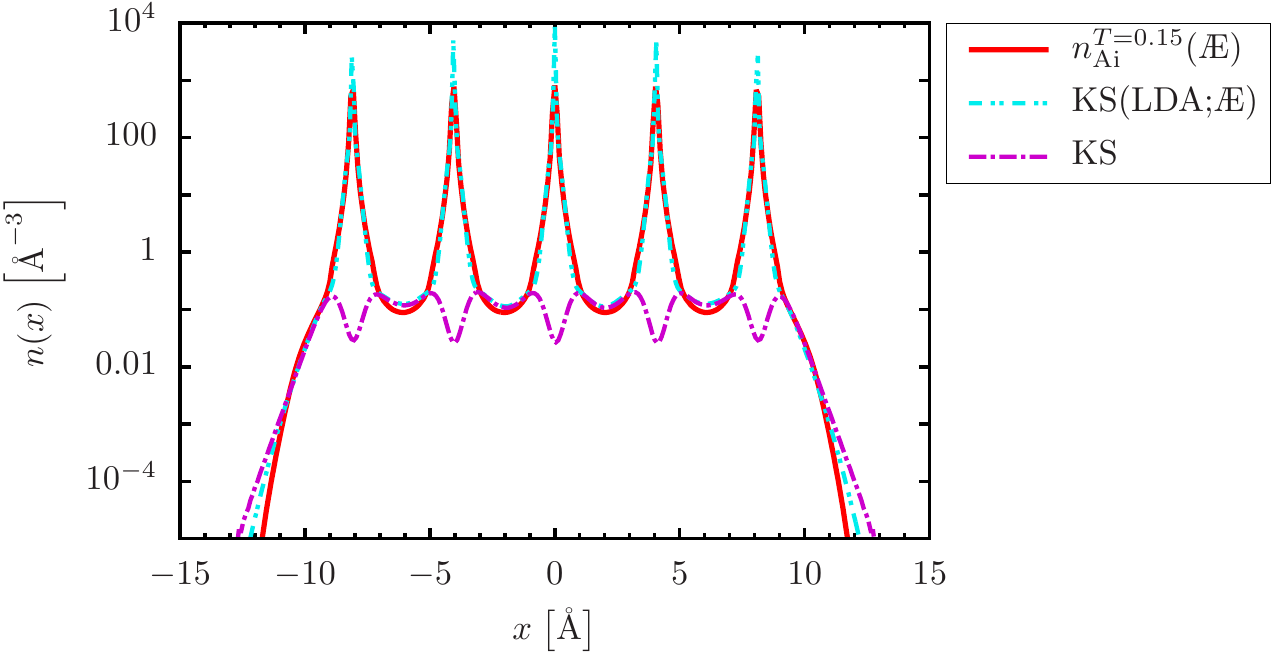}
\caption{\label{201AlnAiT}Comparison of the all-electron density $n_{\mathrm{Ai}}^{T=0.15}(\mbox{\AE})$ with the all-electron KS density (LDA) along the x-axis; see Fig.~\ref{ContourPlots} in Appendix~\ref{NumericalDetails} for $n_{\mathrm{Ai}}^{T=0.15}(\mbox{\AE})$ in the ${z=0}$-plane. The KS valence density (PBE) is the same as that shown in Figs.~\ref{201Aln3pT} and \ref{201Aln3pTlogscale}.}
\end{figure}

The Airy-averaged expression $n_{\mathrm{Ai}}^T$, on the other hand, performs very well exactly where it is designed to do so, namely in the evanescent region across the boundary of classically allowed and forbidden regions. This is evident on the logarithmic scales of Figs.~\ref{201Aln3pTlogscale} and \ref{201AlnAiT}, where we compare KS results with the Airy-averaged valence density and its all-electron version, respectively. Given the variability of KS calculations across different settings, like LDA vs PBE and valence vs all-electron calculations, as exemplified in Figs.~\ref{201Aln3pTlogscale} and \ref{201AlnAiT}, we find $n_{\mathrm{Ai}}^T(\mbox{\AE})$ in Fig.~\ref{201AlnAiT} to match the all-electron KS density reasonably well over eight orders of magnitude, which encompass all parts of the nanoparticle: the electronic structure close to the nuclei, the bond regions, and the exponential decay of the density into vacuum.
  
\FloatBarrier

\section{\label{secConclusions}Conclusions and perspectives}

In this work we extended the semiclassical machinery of density--potential functional theory (DPFT) from low-dimensional settings to three dimensions. Specifically, we developed semiclassical expressions for particle densities and energies of isolated systems that scale (quasi-)linearly with particle number and thus enable, for example, electronic structure calculations of mesoscopic molecules and nanoparticles. Most importantly, those expressions---which we derived in complete analogy to their low-dimensional versions and without relying on ad-hoc measures or assumptions---are void of free parameters and can be further improved in a systematic manner. These characteristics put the semiclassical DPFT formalism that has been developed over the last decade in stark contrast to many contemporary approaches to density functional theory in general and to its orbital-free variants in particular.

To illustrate the generic semiclassical features as well as the practical aspects of our DPFT implementation, we put an emphasis on the calculation of particle densities, also because the improvements of the semiclassical energies upon the Thomas--Fermi model are less convincing, at least for the cases studied here. For the first time DPFT was used to simulate real-world composite atomic systems, specifically, the electronic structure of metal dimers and nanoparticles. As expected, the semiclassical DPFT densities become more accurate as the particle number increases. Indeed, for the 603 (2613) electrons of the valence (all-electron) density of a nanoparticle composed of 201 aluminum atoms, their quality is competitive with that of generic Kohn--Sham calculations.

Our two particle density expressions $n_{\mathrm{Ai}}$ and $n_{3'}$ deliver accurate results in different regimes and for different reasons. The semilocal `Airy-averaged' $n_{\mathrm{Ai}}$ is exact for linear (effective) potentials and is therefore particularly accurate in the evanescent tails toward vacuum. While we may speculate that further systematic improvements are attainable by making a successor of $n_{\mathrm{Ai}}$ exact also for quadratic or higher order polynomials, this route is unexplored at present. In contrast, the common feature of $n_{3'}(\vec r)$ and its successors is their reliance on the (effective) potential in an extended region around the focal position~$\vec r$, which makes them particularly powerful in the bulk. The split-operator-based $n_{3'}$ represents the first step in a ladder of explicit systematic approximations, and two more rungs have already been derived and benchmarked. In fact, the most accurate of these expressions has the same computational complexity as $n_{3'}$, with the proviso that the special function at its heart can eventually be computed efficiently, just like the underlying Bessel function at the core of $n_{3'}$. Also the inaccurate semiclassical energies that are associated with the two approximation schemes employed here are bound to improve once suitable implementations of the higher-order corrections are available.

The increase in accuracy through higher-order DPFT approximations will generally decrease computational efficiency, and problem-specific trade-offs between accuracy and efficiency will have to be made in future applications of semiclassical DPFT. Transferability, however, does not need to be sacrificed along this route. The systematic nature of the here developed functional approximations makes them universally applicable across external potentials and types of interactions.

\acknowledgments

We are grateful to Berthold-Georg Englert for most valuable input and insights. This work has been supported by the National Research Foundation, Singapore and A*STAR under its CQT Bridging Grant and its Quantum Engineering Programme.

\appendix

\section{\label{AppendixAiryST}Derivation of DPFT energies and densities in 3D}

In this appendix we derive the semiclassical potential functionals for the finite-temperature single-particle densities presented in Eqs.~(\ref{n3pF}), (\ref{n3pT}), and (\ref{nAiT}), together with the associated kinetic energies.

\subsection{Suzuki--Trotter-approximated densities and energies}

\textbf{Derivation of $n_{3'}^T$.} We start with deriving Eq.~(\ref{n3pT}) from the finite-temperature version
\begin{align}\label{nT}
n^T(\vec r)=g\bok{\vec r}{\eta_T(\mu-H)}{\vec r}
\end{align}
of Eq.~(\ref{nSTA}). Upon approximating the time-evolution operator by $U_{3'}$ and inverting the Fourier transform of $\eta_T(\,)$ in Eq.~(\ref{nSTA}), we arrive at
\begin{align}\label{nSTAT}
n_{3'}^T(\vec r)=g\int\frac{(\d\vec p_1)(\d\vec p_2)(\d\vec r_1)}{(2\pi\hbar)^{2D}}\,\exp\left(\frac{\I}{\hbar}\vec r_1\cdot(\vec p_1-\vec p_2)\right)\,\eta_T(\mu-H_{3'})\,,
\end{align}
where ${H_{3'}=\big(\vec p_1^2+\vec p_2^2\big)/(4m)+V(\vec r+\vec r_1)}$. With ${\vec p=\hbar\vec k}$ we write
\begin{align}\label{n3pTksecond}
n_{3'}^T(\vec k)&=\mathcal F\big\{n_{3'}^T(\vec r')\big\}(\vec k)=\int(\d\vec r')\,\e{-\I\vec k\vec r'}\,n_{3'}^T(\vec r')\nn\\
&=\frac{g}{(2\pi)^{2D}}\int(\d\vec k_1)(\d\vec k_2)(\d\vec r_1)\,\e{\I\,\vec r_1\cdot(\vec k_1-\vec k_2)}\int(\d\vec r')\,\e{-\I\vec k\vec r'}\eta_T\left(\mu-\frac{\vec k_1^2+\vec k_2^2}{4m/\hbar^2}-V(\vec r'+\vec r_1)\right)
\end{align}
for the Fourier transform of $n_{3'}^T(\vec r)$, where we denote the Fourier transform of a function $f(\vec r)$ as ${f(\vec k)=\mathcal F\{f(\vec r)\}(\vec k)=\int(\d\vec r)\,\e{-\I\vec k\vec r}\,f(\vec r)}$ and implement $\mathcal F$ as a fast Fourier transform using the FFTW library for C++ \cite{Frigo2005}. Defining ${\vec r_2=\vec r'+\vec r_1}$, we express the last integral in Eq.~(\ref{n3pTksecond}) as 
\begin{align}
\e{\I\vec k\vec r_1}\int(\d\vec r_2)\,\e{-\I\vec k\vec r_2}\eta_T\left(\mu-\frac{\vec k_1^2+\vec k_2^2}{4m/\hbar^2}-V(\vec r_2)\right)\,, 
\end{align}
such that
\begin{align}\label{n3pTappendix}
n_{3'}^T(\vec k)&=\frac{g}{(2\pi)^{D}}\int(\d\vec k_1)(\d\vec k_2)\underset{\delta(\vec k+\vec k_1-\vec k_2)}{\underbrace{\int\frac{(\d\vec r_1)}{(2\pi)^{D}}\,\e{\I\,\vec r_1\cdot(\vec k+\vec k_1-\vec k_2)}}}\int(\d\vec r')\,\e{-\I\vec k\vec r'}\eta_T\left(\mu-\frac{\vec k_1^2+\vec k_2^2}{4m/\hbar^2}-V(\vec r')\right)\nn\\
&=\frac{g\,\Omega_D}{(2\pi)^{D}}\int(\d\vec r')\,\e{-\I\vec k\vec r'}\int_0^\infty\d k_1\,k_1^{D-1}\,\eta_T\left(\mu-V(\vec r')-\frac{\hbar^2\vec k^2}{8m}-\frac{\hbar^2\vec k_1^2}{2m}\right)
\end{align}
identifies the Fourier transform of the polylogarithm $\mathrm{Li}_{D/2}$ of order $D/2$ after evaluation of the\linebreak $k_1$-integral:
\begin{align}
n_{3'}^T(\vec k)=-g\left(\frac{m\,k_{\mathrm{B}}T}{2\pi\hbar^2}\right)^{D/2}\int(\d\vec r')\,\e{-\I\vec k\vec r'}\,\mathrm{Li}_{D/2}(z)\,.
\end{align}
Here, ${z=-tb}$, ${t(\vec r')=\e{\sigma}}$, ${\sigma(\vec r')=\big(\mu-V(\vec r')\big)/(k_{\mathrm{B}}T)}$, and ${b(k)=\exp\big(-\hbar^2 k^2/(8m\,k_{\mathrm{B}}T)}\big)$. Since ${D/2>0}$ and $z<0$, we may use the integral representation
\begin{align}
\mathrm{Li}_{D/2}(z)=-\int_0^\infty\d x\frac{tbc}{d+tb}=-\int_0^\infty\d x\int_0^\infty\d y\,t(\vec r')\,\e{-yt(\vec r')}\, c\, \e{-yd/b(k)}\,,
\end{align}
with ${c=x^{D/2-1}/\Gamma[D/2]}$ and ${d=\e{x}}$, thereby disentangling the $\vec r'$-dependence of $\mathrm{Li}_{D/2}(z)$ from its\linebreak $k$-dependence, which allows us to obtain Eq.~(\ref{n3pT}) from the inverse Fourier transform of
\begin{align}\label{n3pTkfinal}
n_{3'}^T(\vec k)=\frac{g}{\Gamma[D/2]}\left(\frac{k_{\mathrm{B}}T}{2\pi{\mathcal U}}\right)^{D/2}\int_0^\infty\d y \int(\d\vec r')\,\e{-\I\vec k\vec r'}\,\underset{f_y(\vec r')}{\underbrace{\e{\sigma-y\,\exp(\sigma)}}}\underset{g_y^D(k)}{\underbrace{\int_0^\infty\d x\,x^{D/2-1}\,\e{-y\exp(x+\kappa)}}}\,.
\end{align}

Both $f_y$ and $g_y^D$, cf.~Eqs.~(\ref{fy}) and (\ref{gyD}), are sufficiently suppressed for ${y\gtrsim40}$. We may thus restrict the support of the $y$-integral to ${0<y<40}$ and calculate it with an adaptive quadrature; we use the Boole rule. For each value of $y$ the inverse fast Fourier transform in Eq.~(\ref{n3pT}) delivers all values on the spatial grid of size $G$ in one go, resulting in a computational cost of $n_{3'}^T(\vec r)$ that scales like ${G\,\log G}$. This contrasts with Eq.~(\ref{n3p}), where the density $n_{3'}(\vec r)$ at each of the $G$ grid points, indexed by $\vec r$, requires a summation over the whole grid. Naturally, there is a trade-off between grid size and accurate enough evaluation of the $y$-integral---as a rule of thumb, $n_{3'}^T$ outperforms $n_{3'}$ for ${G\gtrsim 50^3}$ and $n_{3'}^{\mathcal F}$ (see below) for ${G\gtrsim 100^3}$.

Since $g_y^D$ diverges for ${y\to 0}$, we calculate the $y$-integral from $0$ to ${\epsilon\ll 1}$ separately from the rest of $n_{3'}^T(\vec r)$. For ${y\ll 1}$, we have ${f_y(\vec r')\approx\e{\sigma}(1-y\,\e{\sigma})}$, such that
\begin{align}
\int_0^\epsilon\d y\,\mathcal{F}\left\{f_y(\vec r')\right\}(\vec k)\,g_y^D(k)\approx \mathcal{F}\left\{\e{\sigma(\vec r')}\right\}(\vec k)\int_0^\epsilon\d y\,g_y^D(k)-\mathcal{F}\left\{\e{2\sigma(\vec r')}\right\}(\vec k)\int_0^\epsilon\d y\,y\,g_y^D(k)\,.
\end{align}
Both ${\int_0^\epsilon\d y\,g_y^D(k)}$ and ${\int_0^\epsilon\d y\,y\,g_y^D(k)}$ are bounded from above by ${\e{-\kappa}\Gamma(D/2)}$ (for any ${\epsilon>0}$), and the auxilliary functions $g_y^D(k)$ can be evaluated numerically, tabulated, and interpolated. Starting with any initial guess for $\epsilon$, adequately small values of $\epsilon$ are identified during the adaptive $y$-integration. Alternatively, we may replace the lower integration limit in Eq.~(\ref{n3pTkfinal}) by a small value like ${y=10^{-300}}$.\\

\textbf{Derivation of $n_{3'}^{\mathcal F}$.} For anisotropic situations at ${T=0}$, it is expedient to rephrase the density $n_{3'}$ of Eq.~(\ref{n3p}) in terms of fast Fourier transforms, as done in \cite{Trappe2021b}. The resulting $n_{3'}^{\mathcal F}$ still scales like $G^2$, but the computational cost is reduced by a factor of $\sim$10--40, since only exponentials (not Bessel functions) have to be evaluated: Retracing Eqs.~(\ref{nT})--(\ref{n3pTappendix}) with ${\eta_{T\to0}(\,)=\eta(\,)}$, we get
\begin{align}
n_{3'}(\vec k)&=\frac{g\,\Omega_D}{(2\pi)^{D}}\int(\d\vec r')\,\e{-\I\vec k\vec r'}\int_0^\infty\d k_1\,k_1^{D-1}\,\eta\left(Q^2-\hbar^2\vec k_1^2\right)=\frac{g\,\Omega_D}{(2\pi)^{D}}\int(\d\vec r')\,\e{-\I\vec k\vec r'}\frac{Q^D}{D}\eta\left(Q^2\right)\,,
\end{align}
where ${Q^2=2m\big(\mu-V(\vec r')\big)-\frac{\hbar^2\vec k^2}{4}}$, which results in Eq.~(\ref{n3pF}).

\vspace*{\baselineskip}

\textbf{Kinetic energy.} We obtain approximations of the (ground-state) kinetic energy
\begin{align}
E_{\mathrm{kin}}=-\frac{\hbar^2}{2m}\int(\d\vec r)\,\left(\nab_{\vec r}^2\, n^{(1)}(\vec r;\vec r')\right)_{\vec r'=\vec r}
\end{align}
in terms of the effective potential $V$ by deriving approximate one-body reduced density matrices $n^{(1)}(\vec r;\vec r')$. With the Hamiltonian ${H=T+V}$, Eq.~(\ref{LegendreTF}) becomes
\begin{align}
E_{\mathrm{kin}}&=E_1[V-\mu]-\int(\d\vec r)\,\big(V(\vec r)-\mu\big)\,n(\vec r)\nn\\
&=\tr\{(T+V-\mu)\,\eta(\mu-H)\}+\int(\d\vec r)\,\big(\mu-V(\vec r)\big)\,n(\vec r)\nn\\
&=\tr\{T\,\eta(\mu-H)\}\nn\\
&=-g\int(\d\vec r)(\d\vec r')\,\bok{\vec r}{\left(-\frac{\hbar^2}{2m}\nab_{\vec r}^2\right)}{\vec r'}\bok{\vec r'}{\eta(\mu-H)}{\vec r}\nn\\
&=-g\int(\d\vec r)(\d\vec r')\,\delta(\vec r-\vec r')\,\left(-\frac{\hbar^2}{2m}\nab_{\vec r}^2\right)\bok{\vec r'}{\eta(\mu-H)}{\vec r}\nn\\
&=-g\frac{\hbar^2}{2m}\int(\d\vec r)\,\left(\nab_{\vec r}^2\bok{\vec r}{\eta(\mu-H)}{\vec r'}\right)_{\vec r'=\vec r}\,.
\end{align}
In the spirit of Eq.~(\ref{nSTA}), we thus identify the approximate one-body reduced density matrix
\begin{align}\label{1RDM3p}
n^{(1)}_{3'}(\vec r;\vec r') = g\Int\frac{\d t}{2\pi\I t}\,\e{\frac{\I t}{\hbar}\mu}\,\bok{\vec r}{U_{3'}(t)}{\vec r'} = g\int(\d\vec a)\left(\frac{k_{3'}}{2\pi b}\right)^D J_D(2b\,k_{3'})\,,
\end{align}
which is consistent with the Suzuki--Trotter approximation inherent to $n_{3'}$. Equation~(\ref{1RDM3p}) follows the structure of Eq.~(\ref{n3p}) with ${b=\sqrt{a^2+(\vec a+\vec r-\vec r')^2}}$ and yields Eq.~(\ref{Ekin3p}):
\begin{align}
E_{\mathrm{kin}}^{(3')}=-\frac{\hbar^2}{2m}\int(\d\vec r)\,\left[\nab_{\vec r}^2 n^{(1)}_{3'}(\vec r;\vec r')\right]_{\vec r'=\vec r}=\frac{g\,\Omega_D}{(2\pi\hbar)^D\,(2D+4)\, m}\int(\d\vec r)\,\big[2m\big(\mu-V(\vec r)\big)\big]_+^{\frac{D+2}{2}}\,,
\end{align} 
which can be calculated in lieu of the finite-temperature kinetic energy $E_{\mathrm{kin}}^{(3'),T}$ for small enough $T$.

\subsection{Airy-averaged density and energy}

With the help of Eq.~(\ref{trfA4}) and defining ${F(A=H-\mu)=\mathcal E_T(H-\mu)}$, cf.~Eq.~(\ref{fHmuT}), with derivative ${F'(A)=\frac{\partial F(A)}{\partial V}=f(A)=\eta_T(H-\mu)}$, we write the functional ${E_1[V-\mu]}$ from Eq.~(\ref{tracef}) and its functional derivative as
\begin{align}
E_1[V-\mu]&=\tr\{F(A)\}=g\int\frac{(\d\vec r)(\d\vec p)}{(2\pi\hbar)^3}[F(A)]_W\nn\\
&\cong E_1^{\mathrm{Ai},T}[V-\mu]=\frac{g}{(2\pi\hbar)^3}\int(\d\vec r)\int(\d\vec p)\,{\left\langle\!F\big(\tilde{A}\W\big)-\frac{\hbar^2(\nab^2V)}{12m}F''\big(\tilde{A}\W\big)\!\right\rangle}_{\hspace*{-0.2em}\mathrm{Ai}}
\end{align}
and
\begin{align}
n[V-\mu]&=g\int\frac{(\d\vec p)}{(2\pi\hbar)^3}[f(A)]_W\nn\\
&\cong n_{\mathrm{Ai}}^T[V-\mu]=\frac{g}{(2\pi\hbar)^3}\int(\d\vec p)\,{\left\langle\!f\big(\tilde{A}\W\big)-\frac{\hbar^2(\nab^2V)}{12m}f''\big(\tilde{A}\W\big)\!\right\rangle}_{\hspace*{-0.2em}\mathrm{Ai}}\,,
\end{align}
respectively. With
\begin{align}
\int_0^\infty\d p\,4\pi p^2\,f\big(\tilde{A}\W\big)=-(2\pi m k_{\mathrm{B}}T)^{3/2}\,\mathrm{Li}_{3/2}\left(-\e{-\nu_x(\vec r)}\right)
\end{align}
and, in cylindrical coordinates ${\left\{q=\sqrt{p_x^2+p_y^2},\phi,p_z\right\}}$,
\begin{align}
\int(\d\vec p)\,f''\big(\tilde{A}\W\big)&=\int\d p_z\,\frac{2\pi}{(k_{\mathrm{B}}T)^2}\int_0^\infty\d q\, q\,f''\big(\tilde{A}\W\big)\nn\\
&=\frac{2}{(k_{\mathrm{B}}T)^2}\int_0^\infty\d p_z\,\frac{2\pi m k_{\mathrm{B}}T}{4\,\cosh^2\left(\nu_x/2+p_z^2/(4m k_{\mathrm{B}}T)\right)}\nn\\
&=-\frac{(2\pi m k_{\mathrm{B}}T)^{3/2}}{(k_{\mathrm{B}}T)^2}\,\mathrm{Li}_{-1/2}\left(-\e{-\nu_x(\vec r)}\right)\,,
\end{align}
we get
\begin{align}\label{AppendixnAiT}
n_{\mathrm{Ai}}^T[V-\mu]=-g\frac{(2\pi m k_{\mathrm{B}}T)^{3/2}}{(2\pi\hbar)^3}\int\d x\,\mathrm{Ai}(x)\,\left\{\mathrm{Li}_{3/2}\left(-\e{-\nu_x(\vec r)}\right)-\frac{\hbar^2(\nab^2V)}{12m(k_{\mathrm{B}}T)^2}\mathrm{Li}_{-1/2}\left(-\e{-\nu_x(\vec r)}\right)\right\}
\end{align}
and, with ${\frac{\partial}{\partial V}\mathrm{Li}_{s}\left(-\e{-\nu_x(\vec r)}\right)=-\frac{1}{k_{\mathrm{B}}T}\mathrm{Li}_{s-1}\left(-\e{-\nu_x(\vec r)}\right)}$,
\begin{align}\label{AppendixE1AiT}
E_1^{\mathrm{Ai},T}[V-\mu]=\frac{-g(m k_{\mathrm{B}}T)^{3/2}}{(2\pi)^{3/2}\hbar^3}\int(\d\vec r)\int\d x\,\mathrm{Ai}(x)\left\{-(k_{\mathrm{B}}T)\mathrm{Li}_{5/2}\left(-\e{-\nu_x(\vec r)}\right)+\frac{\hbar^2(\nab^2V)}{12m(k_{\mathrm{B}}T)}\mathrm{Li}_{1/2}\left(-\e{-\nu_x(\vec r)}\right)\right\}.
\end{align}
The TF density at finite temperature, which reads
\begin{align}\label{AppendixnTFT}
n_{\mathrm{TF}}^T[V-\mu]=-g\,\left(\frac{m}{2\pi\hbar^2}\right)^{D/2}(k_{\mathrm{B}}T)^{D/2}\,\mathrm{Li}_{D/2}\left(-\e{-\nu(\vec r)}\right)
\end{align}
in $D$ dimensions, is recovered from Eq.~(\ref{AppendixnAiT}) in the case of constant $V$. Analogously,
\begin{align}\label{AppendixE1TFT}
E_1^{\mathrm{TF},T}[V-\mu]=g\,\left(\frac{m}{2\pi\hbar^2}\right)^{D/2}(k_{\mathrm{B}}T)^{\frac{D+2}{2}}\,\mathrm{Li}_{\frac{D+2}{2}}\left(-\e{-\nu(\vec r)}\right)\,.
\end{align}
We find Eq.~(\ref{nAiT}) from integrating Eq.~(\ref{AppendixnAiT}) by parts and exhibiting the units of energy ($\mathcal E$) and length ($\mathcal L$). Analogously, we reveal the computationally more feasible expression
\begin{align}
E_1^{\mathrm{Ai},T}[V-\mu]=-u_0\times\left\{
\begin{array}{ll}
a(\vec r)\int(\d\vec r)\int\d x\,\mathcal{A}(x)\,\left[u_1\,\mathrm{Li}_{3/2}\Big(-\mathrm{e}^{-\nu_x(\vec r)}\Big)+u_2(\vec r)\,\mathrm{Li}_{-1/2}\Big(-\mathrm{e}^{-\nu_x(\vec r)}\Big)\right] & ,\;a(\vec r)>0\\
(k_{\mathrm{B}}T)\int(\d\vec r)\left[u_1\,\mathrm{Li}_{3/2}\Big(-\mathrm{e}^{-\nu_0(\vec r)}\Big)+u_2(\vec r)\,\mathrm{Li}_{-1/2}\Big(-\mathrm{e}^{-\nu_0(\vec r)}\Big)\right] & ,\;a(\vec r)=0
\end{array}
\right.
\end{align}
for Eq.~(\ref{AppendixE1AiT}), see Eqs.~(\ref{beta0})--(\ref{nux}).

\section{\label{NumericalDetails}Additional details on the numerics and miscellaneous results}

In this appendix, we spell out expedient procedures for the numerical evaluation of the DPFT densities and energies. We also analyze the semiclassical density and energy formulae applied to harmonically confined fermion gases and to the dissociation of hydrogen.

\subsection{Hartree potential}

We implement the Hartree potential as follows: With the Fourier transform $n(\vec k)$ of the spatial density $n(\vec r)$, we write the Hartree energy as
\begin{align}\label{EH}
E_{\mathrm{H}}[n]=\frac{W}{2}\int(\d\vec r)(\d\vec r')\frac{n(\vec r)\,n(\vec r')}{|\vec r-\vec r'|}=\frac{W}{2}\int\frac{(\d\vec k)}{(2\pi)^3}\,4\pi\frac{n(\vec k)\,n(-\vec k)}{k^2}\,,
\end{align}
whose discretized version is ill-defined due to the divergent summand at ${k=0}$. We regularize this singularity by adding and subtracting ${n(\vec k=0)^2\,\e{-k^2}= N^2\,\e{-k^2}}$ in the numerator of Eq.~(\ref{EH}):
\begin{align}
E_{\mathrm{H}}[n]&=\frac{W}{4\pi^2}\left[\underset{k\to0}{\mathrm{lim}}\frac{n(\vec k)\,n(-\vec k)-N^2\,\e{-k^2}}{k^2}+\Delta k^3\left(\sum_{\vec k\not=0}\frac{n(\vec k)\,n(-\vec k)-N^2\,\e{-k^2}}{k^2}\right)+\int(\d\vec k)\,\frac{N^2\,\e{-k^2}}{k^2}\right]\nn\\
&=\frac{W}{4\pi^2}\left[N^2+\Delta k^3\left(\sum_{\vec k\not=0}\frac{n(\vec k)\,n(-\vec k)-N^2\,\e{-k^2}}{k^2}\right)+2\pi^{3/2}N^2\right]\,,\label{EHdiscrete2}
\end{align}
where ${\Delta k=2\pi/B}$. For the equality in Eq.~(\ref{EHdiscrete2}) to hold, we assume that ${\nabla_{\vec k}[n(\vec k)\,n(-\vec k)]}$ decays faster than $2\vec k$ as ${|\vec k|\to0}$. The same procedure regularizes the Hartree potential
\begin{align}
V_{\mathrm{H}}(\vec r)&=W\int(\d\vec r')\frac{n(\vec r')}{|\vec r-\vec r'|}\nn\\
&=4\pi W\left[\mathcal F^{-1}\left\{\left.\frac{n(\vec k)-N\,\e{-k^2}}{k^2}\right|_{\vec k\not=0}+\left.N\right|_{\vec k=0}\right\}(\vec r)+\frac{N}{4\pi^2}\left(\left.\frac{\pi}{r}\mathrm{Erf}(r/2)\right|_{\vec r\not=0}+\left.\sqrt{\pi}\right|_{\vec r=0}\right)\right]\,.
\end{align}
The computational cost of both $E_{\mathrm{H}}[n]$ and $V_{\mathrm{H}}(\vec r)$ scales like ${G\,\log G}$.

\subsection{Regularizations of the DPFT densities and energies}

The approximate semiclassical densities do not possess all the features of the exact density. In particular, $n_{3'}$, $n_{3'}^{\mathcal F}$, and $n_{3'}^T$ can exhibit oscillations around zero in the classically forbidden region. We therefore evaluate exchange--correlation functionals that demand nonnegative densities with $\big[n[V]\big]_+$ instead of $n[V]$. This procedure is only justified if negative densities are small in magnitude, which is usually the case for large enough $N$ and (in case of $n_{3'}^T$ or $n_{\mathrm{Ai}}^T$) temperatures that are large enough while retaining the ground-state character of the density profiles. Furthermore, the PBE functional requires spatial derivatives of the densities, and both $n_{\mathrm{Ai}}^T$ and $E_1^{\mathrm{Ai,T}}$ require the gradient and the Laplacian of the effective potential $V$. If the numerical differentiation introduces instabilities in the selfconsistent DPFT loop for strong interactions and if many iterations are required, we regularize the numerical derivatives in a three-step process: The first and second partial derivatives are obtained via fast Fourier transform after a smooth window-function trimming (of the function to be differentiated) near the boundary (${r>r_B=0.75\times B/2}$) of the (large enough) numerical grid. This is already sufficient for selected systems like the noninteracting hydrogen atom in Fig.~\ref{H}. We further regularize the partial derivatives at $\vec r$ via convolution with a normal distribution centered at $\vec r$ (using a standard deviation of $2a$--$6a$ with lattice constant $a$). We then apply the same regularization to the gradient and the Laplacian assembled from the regularized partial derivatives. This systematic approximation of the derivatives becomes increasingly accurate with rising spatial resolution.\\

\subsection{Details on precision, accuracy, efficiency}

\begin{figure}[ht]
\begin{center}
\includegraphics[width=0.325\linewidth]{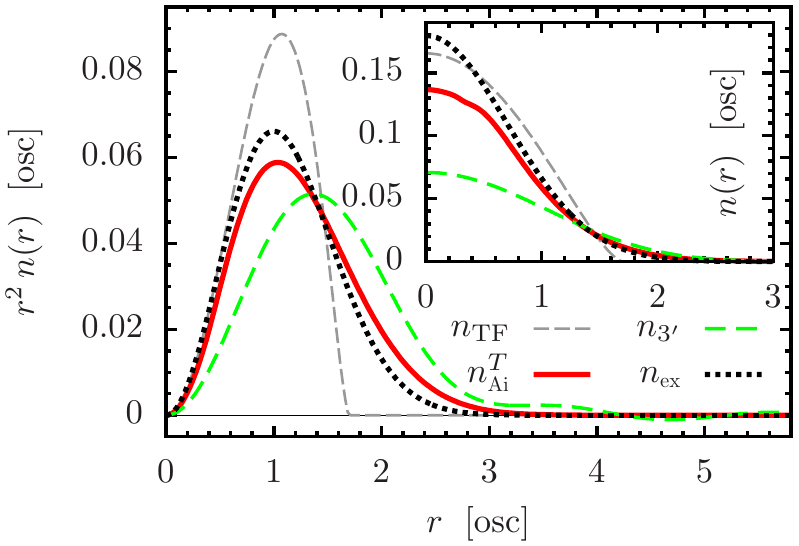}\hfill
\includegraphics[width=0.325\linewidth]{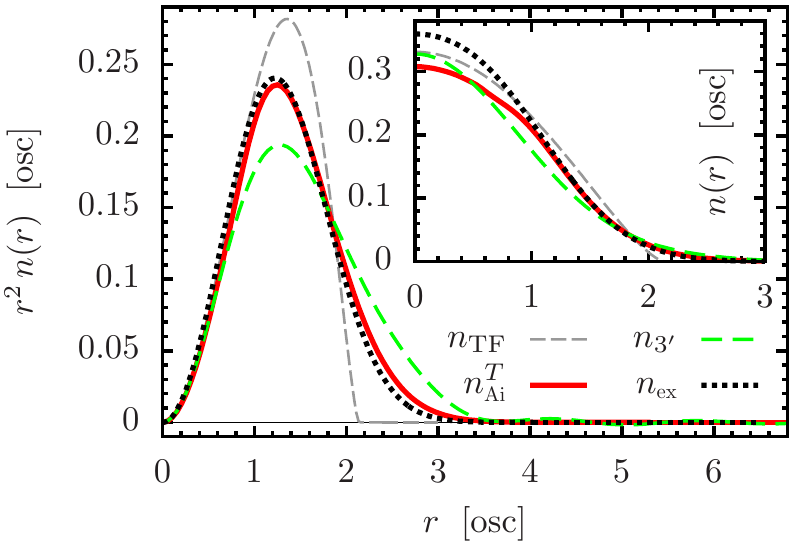}\hfill
\includegraphics[width=0.31\linewidth]{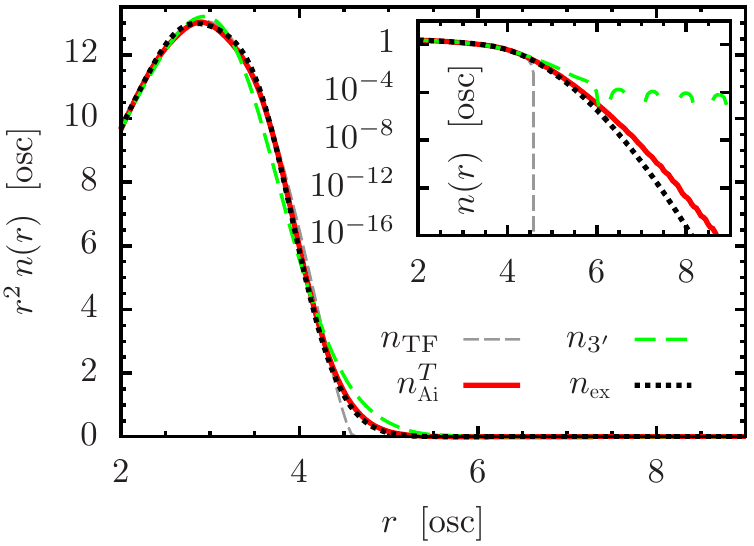}
\caption{\label{HO}Unsurprisingly, the quality of the semiclassical densities improves with larger particle number $N$, here illustrated with the aid of noninteracting unpolarized spin-1/2 fermions in a harmonic trap for ${N=1}$ (left), ${N=4}$ (center), and ${N=400}$ (right). The exact densities for large particle numbers are generated with the (fractionally filled) shell densities reported in Ref.~\cite{Brack2001}.}
\end{center}
\end{figure}

The evaluation of the energy requires a high enough resolution of the integration grid. We obtain an estimate of the required grid size with setups for which we have an exact energy expression in terms of the particle density. In the case of hydrogen, we find a grid of ${\sim400^3}$ points spaced at ${\sim0.026\,}${\AA} sufficient when evaluating the exact energy (in terms of the ground-state density) with the exact density, see Table~\ref{EkinHnonint}. Table~\ref{EkinHnonintSC} reports the selfconsistent kinetic energies based on the semiclassical approximations $n_{3'}$ and $n_{\mathrm{Ai}}^T$ for $384^3$ grid points. In Fig.~\ref{HO} we benchmark our DPFT densities against exact results for the three-dimensional isotropic harmonic oscillator. The TF densities capture the trend of the bulk (reasonably well for ${N=400}$), but are of no use in the classically forbidden region. $n_{3'}$ gives a rough average through the oscillations of the exact density and generates a decay into the classically forbidden region, albeit with too fat a tail and unphysical oscillations around zero. $n_{\mathrm{Ai}}^T$ performs best and yields a reasonable density profile even for ${N=1}$. The exponential decay of its tail for ${N=400}$ is close to the exact behavior---an observation that also holds for smaller $N$.

While the density profiles in Fig.~\ref{HO} based on $n_{3'}$ and, especially, $n_{\mathrm{Ai}}^T$, present an enormous improvement over the TF densities, we gain little over the TF energies, which deviate by ${\sim1\%}$ from the exact values, even as the particle number increases to ${N=400}$, see Table~\ref{EnergiesHO}. The high resolution suggested by the single hydrogen atom for obtaining a precise semiclassical energy makes the energy evaluation for very inhomogeneous systems like nanoparticles a tedious undertaken. Indeed, our calculations do not indicate convergence of energies even when exceeding resolutions of 10 grid points per Angstrom for the 201-atom aluminum nanoparticle. We thus leave the precise and selfconsistent determination of DPFT energies of chemical systems for future study. The integration grids for all computations presented in this work are listed in Table~\ref{IntegrationGrids}. In addition to the challenges in converging the DPFT energies, we also find them not accurate enough for chemistry applications that involve only a few electrons. We demonstrate this in the following by simulating the dissociation of the hydrogen molecule.

\begin{figure}[ht]
\begin{center}
\includegraphics[height=0.3\linewidth]{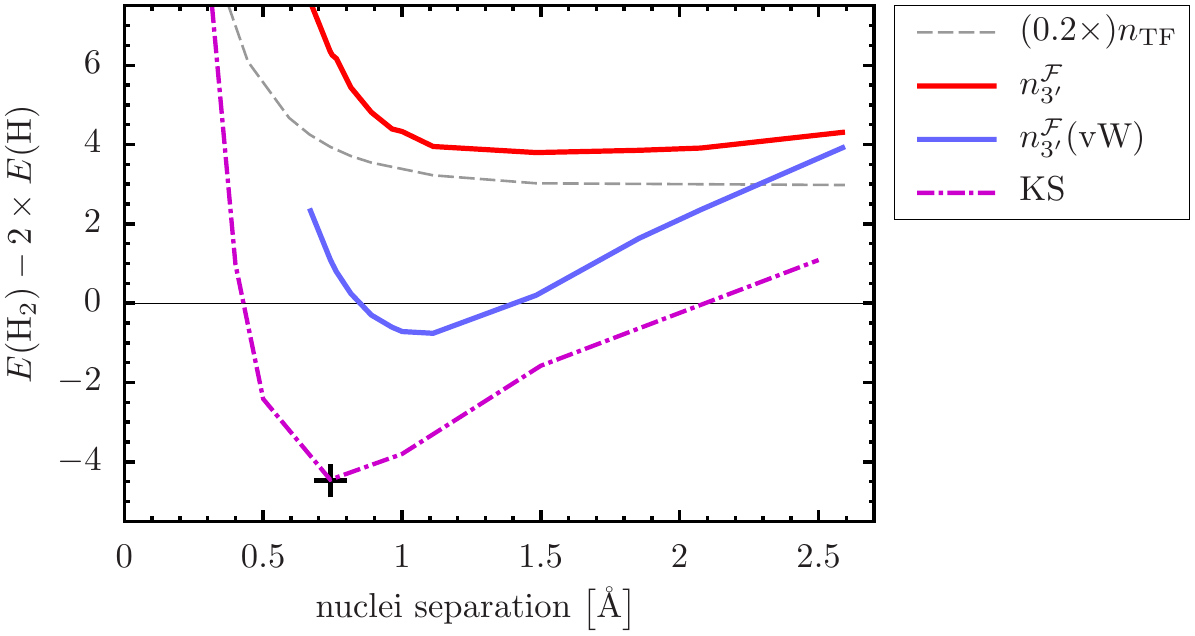}
\caption{\label{H2Dissoziation}$E_{\mathrm{kin}}^{\mathrm{vW}}$, evaluated with $n_{3'}^{\mathcal F}$, yields a qualitatively correct dissociation curve when benchmarked against the selfconsistent KS(PBE) energies. The selfconsistent $n_{3'}^{\mathcal F}$ binds the two hydrogen atoms, albeit weakly, and it is well known that the selfconsistent TF density does not. We thus echo the well-known insight that an accurate kinetic energy is essential for getting the dissociation right. Eventually, however, the selfconsistent DPFT energies associated with the DPFT density formulae deployed in this work are not accurate enough---given today's demands on electronic structure calculations, see also Table~\ref{EkinHnonintSC}.}
\end{center}
\end{figure}

An almost ideal scenario in OF--DFT---maybe second to having the exact universal DFT functional---is to use a kinetic energy expression that is exact when evaluated with the exact density. This is of little value for DPFT calculations, whose defining feature is to avoid using the density functional $E_{\mathrm{kin}}[n]$. However, while we have to live with an approximate interaction energy, we can evaluate the von-Weizs\"acker kinetic energy $E_{\mathrm{kin}}^{\mathrm{vW}}$, which is exact for the H$_2$ singlet ground state, with the converged self-consistent DPFT densities. This serves as a benchmarking exercise for the DPFT energies for small particle numbers $N$. Indeed, in Fig.~\ref{H2Dissoziation} we show that $n_{3'}$ delivers a qualitatively correct picture of the dissociation curve of the hydrogen molecule, which has been troubling generations of DFT practitioners \cite{Cohen2008,Vuckovic2017,Zhang2020}. While $n_{3'}$ binds the two atoms, which improves qualitatively upon the TF model, the accuracy of the dissociation curve is lacking.

\begin{figure}[ht]
\begin{center}
\includegraphics[width=0.345\linewidth]{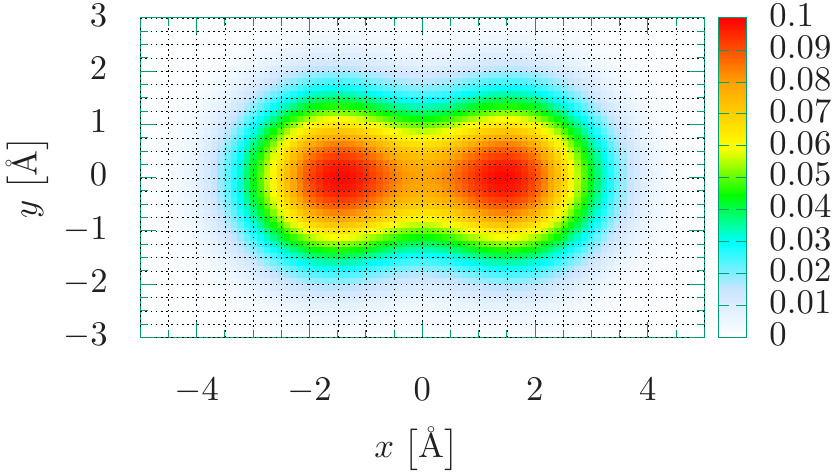}\hfill
\includegraphics[width=0.345\linewidth]{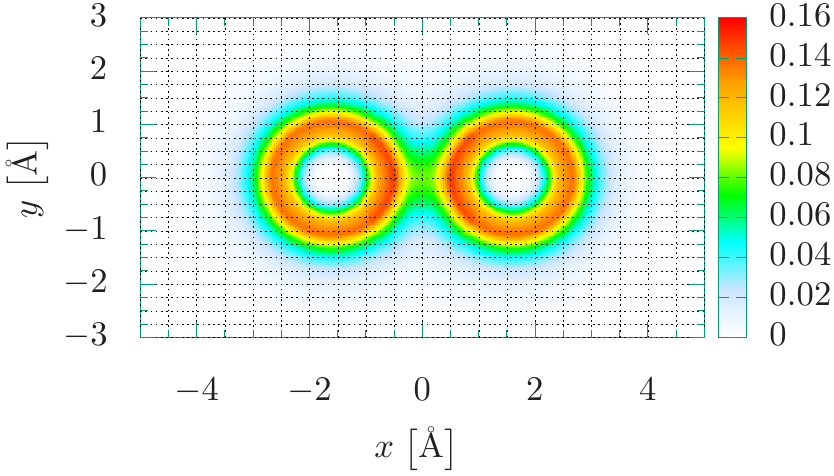}\hfill
\includegraphics[width=0.27\linewidth]{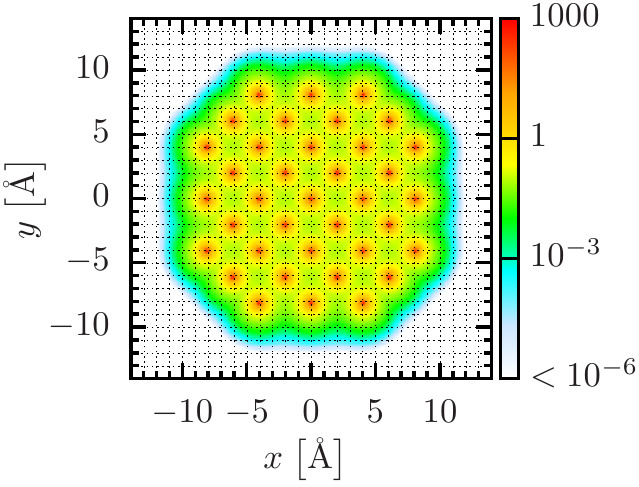}
\caption{\label{ContourPlots} The valence density contours $n_{3'}^{T=0.4}$ (left) and $n_{\mathrm{Ai}}^{T=0.1}$ (center) for the magnesium dimer of Fig.~\ref{pseudoMg2} and the all-electron density contour $n_{\mathrm{Ai}}^{T=0.15}(\mbox{\AE})$ (right) for the aluminum nanoparticle of Fig.~\ref{201AlnAiT}. We display cuts in the plane of ${z=0}$.}
\end{center}
\end{figure}

In Fig.~\ref{ContourPlots} we show contour plots of electronic densities of the Mg-dimer and the Al-nanoparticle illustrated in Figs.~\ref{pseudoMg2} and \ref{201AlnAiT}, respectively.

\begin{table}
\caption{\label{EkinHnonint}An equidistantly spaced integration grid demands a high resolution for extracting high-precision energies. The exact kinetic energy of hydrogen equals the binding energy of ${E_{\mathrm{B}}=13.6057\,}$eV, and can be obtained from the virial theorem since we can evaluate the energy functionals with the known exact density: ${E_{\mathrm{kin}}^{\mathrm{vir}}[n_{\mathrm{ex}}]=-E_{\mathrm{ext}}[n_{\mathrm{ex}}]/2}$. For comparison we report the kinetic energies based on both the exact Coulomb potential and the all-electron pseudopotential of Eq.~(\ref{aePP}). Alternatively, we may enlist the von-Weizs\"acker kinetic energy ${E_{\mathrm{kin}}^{\mathrm{vW}}[n_{\mathrm{ex}}]=9\times E_{\mathrm{kin}}^{\nab^2}[n_{\mathrm{ex}}]}$, which is exact for single-orbital ground states, while the Thomas--Fermi kinetic energy functional ${E_{\mathrm{kin}}^{\mathrm{TF}}[n]}$ is inadequate, even when supplemented with the leading gradient correction $E_{\mathrm{kin}}^{\nab^2}[n]$. All energies are given in eV.}
\begin{ruledtabular}
\setlength\extrarowheight{0.6em}
\begin{tabular}{c|ccccc|c}
&\multicolumn{5}{c}{steps for integration grid}&\\[0.5ex]
functional & 64 & 128 & 256 & 384 & 512 & analytical energy \\[0.5ex]
\hline
$E_{\mathrm{kin}}^{\mathrm{vir}}$ & 13.2171 & 13.5073 & 13.5810 & 13.5947 & 13.5995 & 13.6057 \\[0.5ex]
$E_{\mathrm{kin}}^{\mathrm{vir}}$; Eq.~(\ref{aePP}) & 13.4886 & 13.5891 & 13.6034 & 13.6050 & 13.6054 & --- \\[0.5ex]
$E_{\mathrm{kin}}^{\mathrm{vW}}$ & 13.0525 & 13.5403 & 13.5979 & 13.6034 & 13.6047 & 13.6057 \\[0.5ex]
$E_{\mathrm{kin}}^{\mathrm{TF}}+E_{\mathrm{kin}}^{\nab^2}$ & 9.28932 & 9.37014 & 9.37831 & 9.37902 & 9.37918 & 9.37930 \\[0.5ex]
\end{tabular}
\end{ruledtabular}
\end{table}

\begin{table}
\caption{\label{EkinHnonintSC}The selfconsistent kinetic and total energies of hydrogen from the DPFT densities (cf.~Fig.~\ref{H}) are of no use, although the total energy ${E^{\mathrm{Ai,T}}=E_{\mathrm{kin}}^{\mathrm{Ai,T}}+E_{\mathrm{ext}}\big[n_{\mathrm{Ai}}^{T}\big]}$ for ${k_{\mathrm{B}}T=10^{-6}\,}$eV is not too far off the exact value ${E_{\mathrm{kin}}^{\mathrm{vW}}[n_{\mathrm{ex}}]+E_{\mathrm{ext}}[n_{\mathrm{ex}}]=-13.6057\,}$eV. All energies are given in eV.}
\begin{ruledtabular}
\setlength\extrarowheight{0.6em}
\begin{tabular}{ll|ll|ll}
$E_{\mathrm{kin}}^{\mathrm{TF}}$ & $E^{\mathrm{TF}}$ & $E_{\mathrm{kin}}^{3'}$ & $E^{3'}$ & $E_{\mathrm{kin}}^{\mathrm{Ai,T}}$ & $E^{\mathrm{Ai,T}}$\\
\hline
26.46 & -28.06 & 26.49 & -0.9095 & 40.48 & -12.05 \\ 
\end{tabular}
\end{ruledtabular}
\end{table}

\begin{table}[ht]
\caption{\label{EnergiesHO} The selfconsistent DPFT energies $E$ for spin-unpolarized spin-1/2 fermions in a 3D harmonic trap reaffirm that the semiclassical approximations improve as the particle number increases. The exact energies are ${E_{\mathrm{ex}}(N=4)=8}$ and ${E_{\mathrm{ex}}(N=400)=3210}$ in harmonic oscillator units.}
\begin{ruledtabular}
\setlength\extrarowheight{0.6em}
\begin{tabular}{c|c|c|cc|ccc}
$N$ & density & steps & $E_{\mathrm{kin}}$ & $E_{\mathrm{ext}}$ & $E$ & \parbox{4em}{relative\\ error [\%]} & \parbox{8em}{absolute error\\ per particle $[\hbar\omega]$} \\[0.5ex]
\hline
\multirow{4}{*}{4}&$n_{\mathrm{TF}}$ & 104 & 3.434 & 3.437 & 6.872& 14 & 0.282\\[0.5ex]
&$n_{\mathrm{TF}}$ & 384 & 3.434 &  3.434 & 6.868& 14 & 0.283\\[0.5ex]
&$n_{\mathrm{Ai}}^{T=0.1}$ & 104 & 3.742 &   3.761   & 7.503& 6 & 0.124\\[0.5ex]
&$n_{\mathrm{Ai}}^{T=0.1}$ & 384 &  3.740 &  3.762  & 7.503& 6 & 0.124\\[0.5ex]
\hline
\multirow{7}{*}{400}&$n_{\mathrm{TF}}$ & 104 \& 384 & 1594   & 1594   & 3188& 0.7 & 0.055\\[0.5ex]
&$n_{3'}$ & 104 \& 384 & 1594 & 1636  & 3230& 0.6 & 0.050\\[0.5ex]
&$n_{3'}^{T=1}$ & 104   & 1412   & 1722  & 3134& 2.3 & 0.190\\[0.5ex]
&$n_{3'}^{T=0.1}$ & 104   & 1586   & 1634  & 3220& 0.3 & 0.025\\[0.5ex]
&$n_{\mathrm{Ai}}^{T=0.1}$ & 104 & 1574   & 1615   & 3189& 0.7 & 0.053\\[0.5ex]
&$n_{\mathrm{Ai}}^{T=0.01}$ & 104& 1573   & 1615  & 3188& 0.7 & 0.055\\[0.5ex]
&$n_{\mathrm{Ai}}^{T=0.001}$ & 104  & 1573   & 1615 & 3188& 0.7 & 0.055\\[0.5ex]
\end{tabular}
\end{ruledtabular}
\end{table}   

\begin{table}[ht]
\caption{\label{IntegrationGrids}Details of the numerics pertinent to all the electronic structure results presented in this work: The number $G$ of grid points and the edge length $L$ of the cubic numerical integration grid, as well as the corresponding resolution ${L/\sqrt[3]{G}}$.}
\begin{ruledtabular}
\setlength\extrarowheight{0.6em}
\begin{tabular}{lll|rrr}
Figure & atomic system & density expression & grid points & edge length $\big[\mbox{\AA}\big]$ & resolution $\big[\mbox{\AA}\big]$\\[0.5ex]
\hline
\ref{H} & H & all & $384^3$ & 10 & 0.0260416 \\[0.5ex]
\hline
\ref{Al} & Al & all & $256^3$ & 12 & 0.046875 \\[0.5ex]
\hline
\ref{Mg} & Mg & all & $256^3$ & 16 & 0.0625 \\[0.5ex]
\hline
\multirow{5}{*}{\ref{pseudoMg2}} & \multirow{5}{*}{Mg$_2$} & $n_{\mathrm{TF}}$ & $512^3$ & 10 & 0.0195312 \\[0.5ex]
& & $n_{3'}^T$ & $128^3$ & 15 & 0.1171875 \\[0.5ex]
& & $n_{3'}^{\mathcal{F}}$ & $128^3$ & 15 & 0.1171875 \\[0.5ex]
& & $n_{\mathrm{Ai}}^{T}$ & $512^3$ & 16 & 0.03125 \\[0.5ex]
& & KS & $128^3$ & 14.5 & 0.1132812 \\[0.5ex]
\hline
\multirow{3}{*}{\ref{Al2}} & \multirow{3}{*}{Al$_2$} & $n_{\mathrm{Ai}}^{T}(\mbox{\AE})$ \& KS & $128^3$ & 18 & 0.140625 \\[0.5ex]
& & KS(\AE) & $164^3$ & 14.46421 & 0.0881964 \\[0.5ex]
& & $n_{3'}^T$ & $192^3$ & 18 & 0.09375 \\[0.5ex]
\hline
\multirow{4}{*}{\ref{201Aln3pT}} & \multirow{4}{*}{Al nanoparticle} & $n_{\mathrm{Ai}}^{T}$& $384^3$ & 36 & 0.09375 \\[0.5ex]
& & $n_{3'}^T$ & $256^3$ & 36 & 0.140625 \\[0.5ex]
& & KS(384) & $384^3$ & 36 & 0.09375 \\[0.5ex]
& & KS(256) & $256^3$ & 36 & 0.140625 \\[0.5ex]
\hline
\multirow{3}{*}{\ref{201Aln3pTlogscale}} & \multirow{3}{*}{Al nanoparticle} & $n_{\mathrm{Ai}}^{T}$& $384^3$ & 36 & 0.09375 \\[0.5ex]
& & $n_{3'}^T$ & $256^3$ & 36 & 0.140625 \\[0.5ex]
& & KS & $384^3$ & 36 & 0.09375 \\[0.5ex]
\hline
\multirow{2}{*}{\ref{201AlnAiT}} & \multirow{2}{*}{Al nanoparticle} & $n_{\mathrm{Ai}}^{T}(\mbox{\AE})$ & $384^3$ & 36 & 0.09375 \\[0.5ex]
& & KS(\AE) & $384^3$ & 30 & 0.078125 \\[0.5ex]
\hline
\multirow{3}{*}{\ref{ContourPlots}} & \multirow{2}{*}{Mg$_2$} & $n_{3'}^T$ & $128^3$ & 15 & 0.1171875 \\[0.5ex]
& & $n_{\mathrm{Ai}}^{T}$ & $512^3$ & 16 & 0.03125 \\[0.5ex]
& Al nanoparticle & $n_{\mathrm{Ai}}^{T}(\mbox{\AE})$ & $384^3$ & 36 & 0.09375 \\[0.5ex]
\end{tabular}
\end{ruledtabular}
\end{table} 

\FloatBarrier

\bibliographystyle{bibtexPostdoc}
\bibliography{myPostDocbib}

\end{document}